\documentclass[a4paper,11pt]{article}
\pdfoutput=1 
\usepackage{jcappub} 
\usepackage{aasmacros}
\usepackage{graphicx}
\usepackage{xcolor}
\usepackage{rotating}
\usepackage{amsmath,amssymb,graphics}

\newcommand{\Msun}{M_{\odot}}

\title{Novel constraints on the particle nature of dark matter from stellar streams}

\author[a,b,c]{Nilanjan Banik,}
\author[d]{Jo Bovy,}
\author[b]{Gianfranco Bertone,}
\author[e]{Denis Erkal,}
\author[f]{and T.J.L. de Boer}

\affiliation[a]{Mitchell Institute for Fundamental Physics and Astronomy, Department of Physics and Astronomy, Texas A\&M University, College Station, TX 77843, USA}
\affiliation[b]{GRAPPA Institute, Institute for Theoretical Physics Amsterdam \\
and Delta Institute for Theoretical Physics, University of Amsterdam,Science Park 904, 1098 XH Amsterdam, The Netherlands} 
\affiliation[c]{Lorentz Institute, Leiden University, Niels Bohrweg 2,Leiden, 2333 CA, The Netherlands} 
\affiliation[d]{Department of Astronomy and Astrophysics, University of Toronto, 50 St. George Street, Toronto, ON, M5S 3H4, Canada} 
\affiliation[e]{Department of Physics, University of Surrey, Guildford, GU2 7XH, Surrey, UK UK} 
\affiliation[f]{Institute for Astronomy, University of Hawai`i, 2680 Woodlawn Drive, Honolulu, HI 96822, USA 
}

\emailAdd{banik@tamu.edu}


\abstract{Tidal streams are highly sensitive to perturbations from passing dark matter (DM) subhalos and thus provide a means of measuring their abundance. In a recent paper, we analyzed the distribution of stars along the GD-1 stream with a combination of data from the  {\it Gaia} satellite and  the  Pan-STARRS  survey, and we demonstrated that the population of DM subhalos predicted by the cold dark matter (CDM) paradigm are necessary and sufficient to explain the perturbations observed in the linear density of stars. In this paper, we use the measurements of the subhalo mass function (SHMF) from the GD-1 data combined with a similar analysis of the Pal 5 stream to provide novel constraints on alternative DM scenarios that predict a suppression of the SHMF on scales smaller than the mass of dwarf galaxies, marginalizing over uncertainties in the slope and normalization of the unsuppressed SHMF and the susceptibility of DM subhalos in the inner Milky Way to tidal disruption. In particular, we derive a 95\% lower limit on the mass of warm dark matter (WDM) thermal relics $m_{\rm WDM}>3.6\,\mathrm{keV}$ from streams alone that strengthens to $m_{\rm WDM}>6.2\,\mathrm{keV}$ when adding dwarf satellite counts. Similarly, we constrain the axion mass in ultra-light (``fuzzy'') dark matter (FDM) models to be $m_{\rm FDM}>1.4\times10^{-21}\,\mathrm{eV}$ from streams alone or $m_{\rm FDM}>2.2\times10^{-21}\,\mathrm{eV}$ when adding dwarf satellite counts. Because we make use of simple approximate forms of the streams' SHMF measurement, our analysis is easy to replicate with other alternative DM models that lead to a suppression of the SHMF.}

\begin{document}
\maketitle

\section{Introduction}  

The nature of dark matter (DM) is one of the greatest outstanding mysteries in (astro)physics. For decades now, a simple model consisting of a new particle that behaves as a cold fluid at the onset of structure formation and only interacts significantly through the gravitational force with itself and standard-model matter after its creation in the early Universe, can explain essentially all cosmological and galactic data (e.g., Refs. \cite{Rubin80a,Davis85a,2015PhRvD..92l3516A,Kids2017a,Planck2020a}). The distribution of DM on small scales within galaxies is highly sensitive to the interactions and initial conditions of DM and thus plays a crucial role in efforts to constrain the nature of DM \cite{Weinberg15a,2017ARA&A..55..343B}. In particular, the cold nature of DM causes it to form structure hierarchically starting from small gravitationally-bound halos that merge together to form larger halos \cite{Bond91a} and because the small halos are dense and highly concentrated, many of them survive tidal stripping and should exist as subhalos in galaxies today \cite{Klypin99a,Moore99a}. However, if the initial conditions of DM are warm, as is the case for example in sterile neutrino models \cite{Dodelson94a}, or if DM is an ultra-light axion, then the abundance of these subhalos is significantly reduced \citep{Bode01a,Hui17a}. 

Different probes have been proposed and used in the literature to determine the abundance of DM subhalos as a function of mass in the Milky Way and in external galaxies. Of these, the most sensitive are satellite galaxy counts (e.g., Ref. \cite{Nadler19a}), gravitational lensing (e.g., Ref. \cite{Gilman:2019nap}), and perturbations to stellar streams \cite{Ibata2001,Johnston2002}. We focus on the latter in this paper. Stellar streams are elongated, almost one-dimensional structures produced by the tidal disruption of globular clusters or dwarf galaxies merging into the Milky Way \cite{Johnston1998,Eyre2011,Bovy2014}. The gravitational interactions of the stars in the stream with dark and baryonic substructures can leave a detectable imprint on the stellar density along the stream (e.g., Refs. \cite[]{Ibata2001,Johnston2002,Yoon2011,Carlberg2012,Carlberg2013,Erkal2015,Erkal2015a,Sanders2016,Bovy2016a,Banik2020M}). The extreme fragility of stellar streams means that they are sensititve to DM subhalos down to $10^5\,M_\odot$ with observations possible in the next decade \cite{Bovy2016a}, making them one of the most senstive probes of the nature of DM \cite{Banik2018,Dalal21a}.

In a recently published paper (\citep{Banik2020M}; henceforth B21), we presented a detailed analysis of the distribution of stars in the GD-1 stream, making use of data from {\it Gaia} DR2~\citep{GAIAmain1,GAIAmain2,Lindegren18}, combined with accurate photometry from the Pan-STARRS survey, data release 1~\citep{Chambers16}. By analyzing the power spectrum of density fluctuations in GD-1 as first suggested by Ref. \citep{Bovy2016a}, we have found strong evidence for a population of low-mass DM subhalos in the Galaxy compatible with the predictions of the cold dark matter (CDM) paradigm.

Here, we demonstrate that the evidence for this population of DM substructures sets novel constraints on alternative DM scenarios that predict a suppression of the subhalo mass function (SHMF) on scales smaller than the mass of dwarf galaxies. We derive in particular a stringent lower limit on the mass of thermal warm DM relics (WDM) and on the mass of ultra-light axion DM (fuzzy DM; FDM).

\section{Measurement of the dark-matter subhalo mass function in the inner Milky Way}

In this section, we summarize the techniques used in B21 to constrain the SHMF over the range $[10^6,10^9]\,M_\odot$ using stellar streams. These constraints are then used in the remainder of this paper to constrain alternative DM models. 

The data is described in detail in Ref. \citep{deboer_gd1_2018} and Ref. \citep{Ibata2016} and the modeling framework is described in detail in Refs. \citep{Bovy2016a,Banik2019,Banik2018} and as applied to these data in B21, but we summarize the important aspects of the data and modeling here briefly to allow the reader to understand the basis for our DM substructure measurement and its limitations.

\subsection{The GD-1 stream}

The GD-1 stream was first detected in Sloan Digital Sky Survey (SDSS) data \citep{Grillmair2006}. Subsequent observations with the Canada-France-Hawaii-Telescope (CFHT) revealed a rich density structure with multiple gaps and wiggles in the stream \citep{deboer_gd1_2018}. Recent work with {\it Gaia} revealed $20^{\circ}$ more of the stream \citep{Price-Whelan18a,Webb2018}. Following Ref. \citep{deBoer20a}, we identified a robust sample of GD-1 member stars spanning $60^{\circ}$ on the sky, and presented in B21 the linear density of stars along the stream, excluding stars in the `spur' and `blob' features identified in \citep{Price-Whelan18a} to focus on the thin region of the stream that is most sensitive to perturbations from passing DM subhalos.

In B21, we modeled the GD-1 stream according to the frequency-angle ($\Omega,\theta$) framework introduced by one of us in Ref. \citep{Bovy2014}. We generated and evolved mock streams 
assuming a constant stripping rate from the progenitor cluster throughout its orbit \footnote{We made use of the Python package for galactic dynamics \texttt{galpy}; \url{https://github.com/jobovy/galpy} \citep{Bovy2015}}. Episodic stripping can in principle lead to small stellar density perturbations \citep{Kuepper2009,Kuepper2011}, but this is unlikely to affect the stream away from the progenitor where DM subhalo impacts are expected to leave a detectable signature \citep{Ngan2013}. 

We assume that the current location of the GD-1 progenitor's remnant corresponds to the underdense region located at $\phi_{1} = -40^{\circ}$ in the spherical coordinate system aligned with the stream introduced by~\citep{Koposov2010}. Although the actual location is currently uncertain, a progenitor at $\phi_{1} = -40^{\circ}$ that dissolved $\sim 500$ Myr ago during the previous perigalactic pass is a very likely scenario as shown by the suite of collisional $N$-body simulations of GD-1's disruption in Ref. \citep{Webb2018}. To guard against possible effects of episodic stripping, we conservatively exclude a region of $12^\circ$ around the putative progenitor remnant. In the appendix of B21, we show that assuming that the remnant of the GD-1 progenitor is instead located at $\phi_{1} = -20^{\circ}$ does not change the inferred DM-subhalo abundance. This is essentially the case because the level of perturbations is similar all along the well-populated part of the GD-1 stream and thus it does not matter where we place the progenitor. We also demonstrated that decreasing the size of the cut around the progenitor does not affect the results. We assume a fiducial dynamical age of 3.4 Gyr, again following the detailed simulations of Ref.~\citep{Webb2018}, but we marginalize over stream ages of [3,4,5,6,7] Gyr using a uniform prior when deriving constraints on the SHMF. 

\subsection{The Pal 5 stream}

The Pal 5 stream originates from the tidal stripping of the Palomar 5 globular cluster, located at a sky position of $(\alpha,\delta) \approx (229^\circ, -0^\circ\!.11)$ and at a distance of $\approx$ 23 kpc \citep{Odenkirchen:2003ga}. The advantage of the Pal 5 stream is that its progenitor system is known and there is therefore no uncertainty due to the unknown progenitor location. One disadvantage of the Pal 5 stream is that it is more than twice as far away as GD-1, making it difficult to detect with \emph{Gaia} data \cite{Starkman2020} and \emph{Gaia} data cannot be easily used to help determine the density fluctuations in the stream. Instead, we rely on the density measurements from Ref. \cite{Ibata2016}. A further disadvantage of the Pal 5 stream is its proximity to the Galactic disk and its prograde orbit with an apocenter of $\approx 8\,\mathrm{kpc}$; these properties result in the Pal 5 stream being strongly perturbed by the baryonic substructures such as the bar \citep{Erkal2017,Pearson2017,Banik2019}, the spiral arms \citep{Banik2019} and the giant molecular clouds (GMCs) \citep{Amorisco2016,Banik2019}. In particular, the bar and GMCs likely perturb the Pal 5 stream as much or more than DM substructure with the abundance expected in the standard CDM paradigm. However, observations of the Pal 5 stream are still useful, because they place an upper limit on the contribution from DM subhalos.

We adopt Pal 5 stream density data from Ref. \citep{Ibata2016} for the trailing arm (the leading arm is only detected over a few degree) and apply the same treatment as in Ref. \citep{Bovy2016a}. Ref. \citep{Ibata2016} employs $g$ and $r$-band data obtained using CFHT to determine star counts near the Pal 5 stream down to $g = 24$. Using a simple color-magnitude filter centered on the main-sequence of the Pal 5 cluster, they derive star counts along the stream over the range $20 < g < 23$ and we use the data shown in their Figure 7. Because the data are not corrected for back- and foreground stars, we use the CFHT photometric data to estimate
a constant background level (in good agreement with the more sophisticated analysis from Ref. \cite{Erkal17a}) and subtract this from the density data. 

To remove large-scale trends in the density profile due to orbital phase variations along the stream, possible unmodeled background variations, and unmodeled variations in the stripping rate, we fit the density profiles of both the GD-1 and Pal 5 streams with a third order polynomial and divide the data by this. The resulting relative density variations are the data that we model with perturbations from baryonic and dark substructures. As was done in Ref. \citep{Bovy2016a}, we checked that using either a second or fourth order polynomial gives consistent results as the fiducial third-order polynomial, demonstrating that the exact modeling of the large-scale variations is unimportant for our derived constraints on the nature of DM.

\subsection{Baryonic substructures}

Following Ref. \citep{Banik2019}, we evolve mock GD-1 and Pal 5 streams in the Milky Way potential taking into account perturbations from the known baryonic substructures such as the Galactic bar, the spiral arms, GMCs, globular clusters (GCs), as well as (as we discuss in the next section) DM substructures. We briefly summarize here the implementation of baryonic effects, and refer the reader to B21 for further details:

\begin{itemize}
\item \textit{Bar}---We model the bar with a triaxial, exponential density profile \citep{Wang2012}. We assume a mass of $10^{10}~ \rm{M}_{\odot}$, a pattern speed of 39 km s$^{-1}$kpc$^{-1}$ \citep{Portail2016,Bovy:2019uyb}, an age of 5 Gyr, and an angle between the bar's major axis and our line-of-sight to the Galactic center of 27$^{\circ}$ \citep{Wegg2013}. 
\item \textit{Spiral arms}---We model spiral arms using the model from Ref. \citep{Cox2002}, assuming four arms,  a pattern speed of 19.5 km s$^{-1}$kpc$^{-1}$, radial scale length of 3 kpc, and vertical scale height of 0.3 kpc. Because GD-1 is on a retrograde orbit, the density perturbations induced from the bar and spiral arms turn out to be negligible, but they strongly affect the Pal 5 stream. 
\item \textit{GMCs}---We draw the position and velocity of GMCs with mass larger than $10^{5}$ M$_{\odot}$ from Ref. \citep{Miville-Deschenes2016}, correcting for empty patches on the other side of the Galactic center, and producing multiple realizations by adding random rotations as in Ref. \citep{Banik2019}. We set GMCs on circular orbits and model their internal mass distribution as a Plummer sphere with a scale radius equal to 1/3 of their observed radius. 
\item \textit{GCs}---We adopt the sky and phase space coordinates, mass, and size information of 150 GCs from \citep{Vasiliev2018}. In order to explore the range of possible orbits of the GCs and their effects on the stream density, we draw proper motions and line of sight velocities from a Gaussian centered around the mean value and width equal to the observational uncertainties. As in the case of GMCs, we model GCs as Plummer spheres. GCs do not significantly perturb either the GD-1 or Pal 5 streams.
\end{itemize}

\subsection{Dark matter substructures}\label{sec:dm}

\label{sec:subhalos}

\begin{figure*}
\includegraphics[width=1\textwidth]{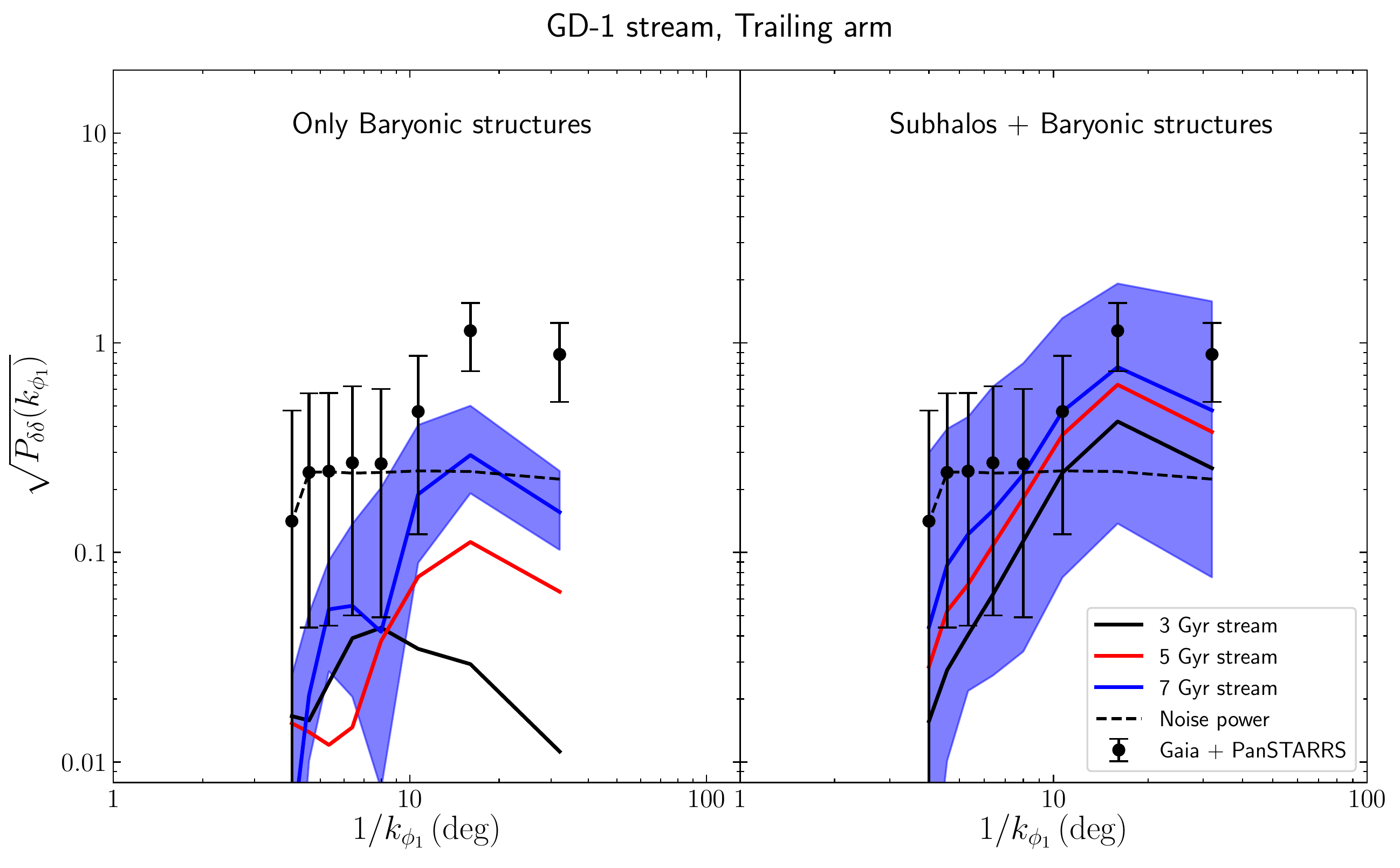}
\caption{Density power spectrum of the trailing arm of the GD-1 stream from B21 compared with that of mock GD-1 streams of different ages perturbed only by baryonic substructures (left panel) and by baryonic plus DM substructures (right). The data points show the median power of the observed stream (black dots) and the $2\sigma$ scatter due to noise in the observed density (error bars). The black dashed line represents the median power of the noise in the density. The solid lines represent the median density power of 1,000 mock streams with age 7 Gyr (blue line) 5 Gyr (red) and 3 Gyr (black). The blue shaded region shows the $2\sigma$ scatter of the density power for the 7 Gyr old stream. CDM substructures are necessary and sufficient to explain the observed density fluctuations.}
\label{fig:Pk_GD-1}
\end{figure*}

Following our previous works \citep{Bovy2016a,Erkal2016,Banik2018}, we consider populations of subhalos with a mass function that follows a standard CDM SHMF in the mass range $[10^{5} - 10^{9}]\,\Msun$. For our fiducial CDM model, we adopt a mass function $dN/dM \propto M^{-1.9}$ and a radial distribution inside the Milky Way that follows an Einasto profile~\citep{Springel2008}. The resulting normalized subhalo profile for the Milky Way is~\citep{Erkal2016}
\begin{equation}
\left(\frac{dn}{dM}\right)_{\rm{CDM}} = c_0\times2.02 \times 10^{-13}~\Msun^{-1}~\mathrm{kpc}^{-3}\left(\frac{M}{m_{0}}\right)^{-1.9}\exp\left\{ - \frac{2}{\alpha_{r}}\left [\left(\frac{r}{r_{-2}}\right)^{\alpha_{r}} - 1 \right]\right\}
\label{eq:dndMc}
\end{equation}
where $c_0=1$, $m_{0}= 2.52\times 10^{7}~\Msun$, $\alpha_{r} = 0.678$ and $r_{-2} = 162.4$~kpc. We use this fiducial mass function to obtain the fiducial number density of subhalos in the ranges $[10^6,10^7]\,M_\odot$, $[10^7,10^8]\,M_\odot$, and $[10^8,10^9]\,M_\odot$. To measure the SHMF, we then allow these three number densities to vary independently from each other around their fiducial values to obtain a measurement of the SHMF in these three bins. Within each bin, we assume that the mass function is $dN/dM \propto M^{-1.9}$, but the bins are narrow enough that this does not affect our results. As discussed in B21, we have also performed full stream analyses using WDM models and their results are virtually indistinguishable from the WDM constraints that we derive below from the subhalo measurements in the three mass bins.

As discussed in detail in B21, we ignore the time evolution of the number density of the subhalo population, and the disruption of subhalos due to the Milky Way disk and other baryonic effects in the measurement. However, we will account for tidal disruption in our further modeling below. We model the velocity distribution of DM subhalos as an isotropic Gaussian with a one-dimensional velocity dispersion of $120\,\mathrm{km\,s}^{-1}$. We describe the subhalos as Plummer spheres with scale radius $r_{s} = 1.62\ \rm{kpc} \ (M_{\rm{sub}}/10^{8}\Msun)^{0.5}$ \citep{Erkal2016}. We stress that at large angular scales, where the signal dominates noise, the density power spectrum induced by subhalos with a Hernquist profile is identical to that of subhalos with a Plummer profile~\citep{Bovy2016a}; at these large scales, the internal structure of the subhalos also does not influence the predicted stream perturbations \citep{Bovy2016a}, so any changes to the concentration-mass relation in alternative DM models are unimportant for the SHMF measurement.  Random sets of impacts are then sampled along the past orbit of the stream out to five times the scale radius of the subhalos and their effect on the stream density is computed using the impulse approximation using the fast frequency-angle method first described in Ref.~\cite{Bovy2016a}. In that paper, extensive tests of the framework against $N$-body simulations are described that demonstrate that it precisely predicts the perturbation density power spectrum for a given population of DM subhalos.

\subsection{Power spectrum analysis} 

Starting from the normalised linear density along the stream angle $\phi_1$ for each stream, we compute the 1-dimensional power spectrum of the density contrast $P_{\delta\delta}(k_{\phi_1}) = \langle \delta(k_{\phi_1})^2 \rangle$. For GD-1, we show in Fig. \ref{fig:Pk_GD-1} the square root of the power $\sqrt{P_{\delta\delta}(k_{\phi_1})}$ as a function of the inverse of the frequency (1/$k_{\phi_1}$), which encodes the correlation in density contrast as a function of the angular scale. The black points with error bars correspond to the observed power spectrum in GD-1 data. The dashed line shows the median power of the density noise. As one can see, there is a clear excess above the noise floor in the last two bins, which correspond to angular scales larger than 10 degrees. 

We compare the observed power spectrum with the one predicted with mock realizations of baryonic substructures (left panel) and baryonic plus DM substructures (right panel). The blue, red and black solid lines show the median density power of 1000 mock GD-1 stream realizations that are 7, 5 and 3 Gyr old respectively. As expected, older mock streams exhibit larger power on almost all length scales, as they have more time to interact with substructures and consequently gets more perturbed by them. Even assuming a stream as old as 7 Gyr, however, baryonic structures alone cannot account for the observed density power of the stream. When CDM substructures are included, mock streams exhibit a power spectrum fully consistent with the data. We stress that the curves in the right panel of Fig. \ref{fig:Pk_GD-1} are not a fit to the data, but an actual prediction based on a `vanilla' description of CDM substructures in the Milky Way. We thus conclude---in the context of particle DM models---that CDM substructures are necessary and sufficient to explain the density fluctuations in the GD-1 stream.

\subsection{Sufficient summaries of the Pal 5 and GD-1 stream analyses}\label{sec-summaries}

In B21, we used the formalism described in this section to fully forward model stream density fluctuations of the Pal 5 and GD-1 streams in the presence of baryonic substructure and DM subhalos in a variety of models. For example, in the context of CDM, we determined the amplitude of the SHMF for a standard CDM power-law mass-function by directly comparing models with different mass-function amplitudes to the stream density data. However, such an analysis is expensive and difficult to repeat for different DM models. To be able to easily constrain different DM models, we use here the intermediate level summary of the B21 constraints on the abundance of DM subhalos that is their abundance in two mass ranges where the abundance is well constrained: $[10^7,10^8]\,M_\odot$ and $[10^8,10^9]\,M_\odot$. In the range $[10^6,10^7]\,M_\odot$, B21 only found an upper limit that is not very informative and we do not use here, although we do include it for reference in Figs. \ref{fig:subhalo_mass_function}, \ref{fig:subhalo_mass_function_fdm}, and \ref{fig:subhalo_mass_function_altdm}. 

As discussed in B21, the posterior distribution functions from combining both streams for the subhalo abundance in these mass ranges relative to that predicted by the fiducial CDM model from Equation \eqref{eq:dndMc}, can be approximated as follows: for $[10^7,10^8]\,M_\odot$
\begin{align}
    \ln p(r=\log_{10}[n_\mathrm{sub}/n_{\mathrm{sub, CDM}}]) = & -\frac{|r+0.5|^{2.5}}{2\times 0.5^2}\,,
\end{align}
and for $[10^8,10^9]\,M_\odot$
\begin{align}
    \ln p(r=& \log_{10}[n_\mathrm{sub}/n_{\mathrm{sub, CDM}}]) \nonumber\\ 
    & = -\frac{|r+0.7|^{2}}{2\times 0.3^2}\qquad (r < -0.7)\\
    & = -\frac{|r+0.7|^{2}}{2\times 0.6^2}\qquad (r \geq -0.7)\,.
\end{align}
Constraints on the mass of a thermal relic WDM particle from just using these two measurements are almost exactly the same as those derived using full forward modeling of the streams data. Therefore, for the purpose of constraining the nature of DM using the observed SHMF, these two constraints are essentially sufficient statistics of the full streams data. Using only these two measurements, it is easy to repeat the analysis in this paper for other models for the WDM mass function and for other DM models. Therefore, we only rely on these two streams SHMF measurements in this paper.

\section{Alternative dark matter models}\label{sec:altdm}

We consider two popular alternative DM models: warm and fuzzy DM. In both of these models, the SHMF can be written as a suppression of the fiducial CDM mass function from Equation \eqref{eq:dndMc} in a way that only depends on a single parameter, which in both cases is the mass of the DM particle. For WDM, the suppression is a simple multiplicative factor, but in the case of FDM, there is an additional term that does not multiply the CDM mass function. Thus, in both cases, we can write the SHMF as
\begin{equation}
\left(\frac{dn}{dM} \right) =  \mathcal{S}(M;m_\mathrm{DM})\,\left(\frac{dn}{dM} \right)_{\rm{CDM}}+\mathcal{T}(M;m_\mathrm{DM})\,,
\label{eq:altdmcdm}
\end{equation}
where $\mathcal{T}(M;m_\mathrm{DM}) = 0$ for WDM.

The mass functions obtained using standard prescriptions for $\mathcal{S}(M;m_\mathrm{DM})$ and $\mathcal{T}(M;m_\mathrm{DM})$ do not account for the tidal disruption of subhalos by the presence of a baryonic disk and bulge in the Milky Way. The presence of the baryonic components in the Milky Way leads to additional disruption of DM subhalos and, thus, a lowering of the expected subhalo abundance in all DM models \citep{DOnghia2010,Sawala2016,GarrisonKimmel19a,Webb2020}. This subhalo depletion is highest for DM subhalos with orbits that bring them close to the disk and, therefore, the expected depletion varies with radius, from little depletion at radii beyond 100 kpc and an increasing depletion as one approaches the center. To account for this, we multiply the mass function $dn/dM$ of subhalos in the inner Milky Way where the streams are orbiting by a factor $f_\mathrm{survive}$. 

All of the models below therefore have four parameters: the amplitude $c_0$ of the unsuppressed mass function from Equation \eqref{eq:dndMc} and its logarithmic slope $\alpha$, the fraction $f_\mathrm{survive}$ of subhalos in the inner Milky Way that survive tidal disruption by the baryonic component, and the DM mass $m_{\mathrm{DM}}$. We allow the overall subhalo normalization $c_0$ to vary by a factor of 100 smaller and larger than the fiducial value given below Equation \eqref{eq:dndMc} and $f_\mathrm{survive}$ between $0.01\%$ and $50\%$; in both cases we use logarithmic priors. To account for uncertainty in the logarithmic slope of the CDM mass function \citep{Springel2008}, we allow $\alpha$ to vary in the range $[-1.95,-1.85]$ with a uniform prior. We consider various priors for the DM mass discussed below.

\subsection{Warm dark matter}\label{sec:wdm}

We model WDM as a thermal relic whose properties are determined by its particle mass. Similar to other subgalactic constraints on the mass of thermal relic WDM \cite{Gilman:2019nap,Nadler21a}, we adopt the mass function from Ref. \citep{Lovell2013} that provides an accurate description of the WDM subhalos population of a Milky-Way-like galaxy observed in the Aquarius simulation \citep{Springel2008} (cf. Equation \ref{eq:altdmcdm})
\begin{equation}
\left(\frac{dn}{dM} \right)_{\rm{WDM}} =  \left(1+ \gamma\frac{M_{\rm{hm}}}{M} \right)^{-\beta} \left(\frac{dn}{dM} \right)_{\rm{CDM}},
\label{eq:wdmcdm}
\end{equation}
where $\gamma = 2.7$ and $\beta = 0.99$. The half-mode mass $M_{\rm{hm}}$, which parameterises the subhalo mass below which the mass function is strongly suppressed, is given by
\begin{equation}
    M_{\rm{hm}} = {4\pi \Omega_m \rho_{\mathrm{crit}} \over 3}\,\left(\lambda_{\rm{hm}}\over 2\right)^3
\end{equation}
and is thus equal to the mean mass contained within a radius of half-mode wavelength $\lambda_{\rm{hm}} = 2\pi \alpha_{\rm{cutoff}}(2^{\nu/5} - 1 )^{-1/2\nu}$, with $\nu = 1.12$ \citep{Viel2005} where $\Omega_m$ is the matter density parameter, $\rho_{\mathrm{crit}}$ is the critical density, and
\begin{equation}
\alpha_{\rm{cutoff}} = 0.047\left(\frac{m_{\rm{WDM}}}{\rm{keV}}\right)^{-1.11}\left(\frac{\Omega_c}{0.2589}\right)^{0.11} 
 \left(\frac{h}{0.6774}\right)^{1.22}h^{-1}\rm{Mpc},
\end{equation}
where $m_{\rm{WDM}}$ is the WDM particle mass and $\Omega_c$ is DM density parameter \cite{Schneider12a}. We have assumed here that all of the DM is warm, i.e., $\Omega_{\rm{WDM}} = \Omega_c$

Finding and resolving low-mass subhalos in $N$-body simulations is complex \cite{Onions12a,vbd18a}, especially in WDM simulations, and the abundance of WDM subhalos for a given WDM mass is therefore still debated. Recently, Ref. \cite{Lovell21a} has suggested that the effect of tidal stripping of WDM subhalos is such that the abundance of WDM subhalos in the inner $\approx 40\,\mathrm{kpc}$ is significantly less suppressed than Equation \eqref{eq:wdmcdm} with $\gamma = 2.7$ and $\beta = 0.99$ would suggest. In particular, for $m_{\rm{WDM}} = 3.3\,\mathrm{keV}$ they find that subhalos in the range $[10^7,10^8]\,M_\odot$ are three times more abundant, while those in the range $[10^8,10^9]\,M_\odot$ are 1.75 times more abundant than Equation \eqref{eq:wdmcdm} would predict, albeit with strong halo-to-halo scatter. Using the form of Equation \eqref{eq:wdmcdm}, we find that setting $\gamma = 1.5$ and $\beta = 0.7$ approximately matches these modified abundances for $m_{\rm{WDM}} = 3.3\,\mathrm{keV}$. To explore the effect of a less strong suppression in the inner Milky Way, below we perform fits with these values of $\beta$ and $\gamma$ as well.

\subsection{Fuzzy dark matter}

Detailed numerical simulations of the SHMF in FDM models have not yet been performed because of the difficulty in achieving the necessary resolution to resolve the quantum effects near the de Broglie wavelength in a cosmological simulation. In the absence of these, we follow the same approach as taken by Ref. \cite{Schutz20a} and use the results from extended Press-Schechter calculations combined with prescriptions for tidal stripping of DM subhalos. These calculations were performed in Ref. \cite{Du17a} using \texttt{galacticus} \cite{galacticus}. The resulting SHMF is given by
\begin{equation}
\begin{split}
\left(\frac{dn}{dM} \right)_{\rm{FDM}} = &  \left[1+\left({M \over 2 \,m_{22}^{-1.6}\times 10^8\,M_\odot}\right)^{-0.72}\right]^{{-13.9}}\,\left(\frac{dn}{dM} \right)_{\rm{CDM}}\\& +{0.014\,m_{22}^{1.5} \over M}\exp\left[-\left(\ln\left\{{M \over 4.7\,m_{22}^{-1.5}\times 10^8\,M_\odot}\right\}\right)^2/1.4\right]\,,
\label{eq:fdmcdm}
\end{split}
\end{equation}
where $m_{22} = m_{\mathrm{FDM}} / 10^{-22}\,\mathrm{eV}$. Compared to WDM, the cut-off in the FDM SHMF is much sharper, essentially because the transfer function for FDM has a much sharper cut-off than the WDM transfer function. DM subhalos are the only form of DM substructure that we assume in FDM models for our analysis here. That means that we ignore the density fluctuations arising from wave interference in the evolving DM distribution on scales of the de Broglie wavelength $\lambda \approx 600\,\mathrm{pc}/m_{22}$ \cite{Hui17a}. These density fluctuations themselves perturb tidal streams in a way that mimics CDM \cite{Dalal21a}. However, we will see that our constraints are at $m_{22} \approx 20$, where the density fluctuations are far less important \cite{Amorisco18a}, such that we can ignore them and focus on the SHMF.

\section{Constraints on warm dark matter}

\subsection{Constraints from streams alone}\label{sec:wdm_wstreams}

\begin{figure}
\centering
\includegraphics[width=0.8\textwidth]{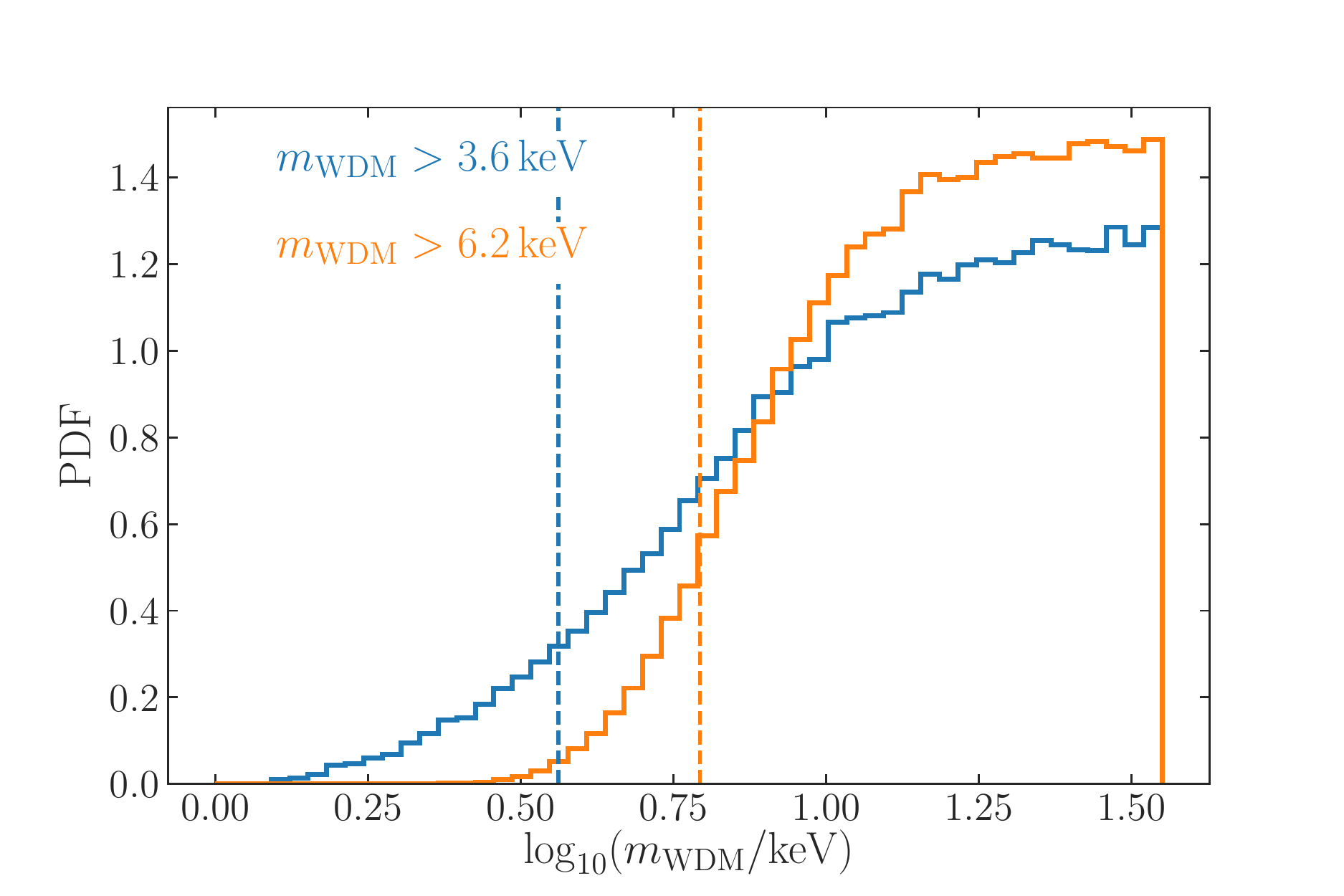}
\caption{Posterior PDF for the thermal WDM particle mass using density fluctuations in the Pal 5 and GD-1 streams alone (blue curves) and combining this with classical-satellite counts (orange curves). The 95\% lower bounds of both PDFs are indicated with the labels and dashed vertical lines.}
\label{fig:mwdm_PDF}
\end{figure}

The evidence for CDM substructures obtained in B21 can be used to constrain any particle DM that leads to a substantial suppression of the DM SHMF on sub-dwarf scales, and thus of the linear density contrast power spectrum of streams. We start here with the case of thermal WDM candidates. Following Ref. \cite{Gilman:2019nap}, we use a uniform prior on $\log_{10} M_{\mathrm{hm}} / M_\odot$ over the range $[4.8,10]$.  This corresponds to a range of $m_\mathrm{WDM}$ of approximately 1 to 50 keV. The lower bound of $M_{\mathrm{hm}}  = 10^{4.8}\,M_\odot$ is well below the sensitivity of our stream analysis method. Here, we use only the subhalo abundance constraints derived from the streams data alone and we fit the entire model for the SHMF that varies the slope $\alpha$ over the range $[-1.95,-1.85]$, the amplitude $c_0$ and $f_\mathrm{survive}$ within their wide ranges discussed in Section \ref{sec:altdm}, and the warm DM mass. Because we only use streams in the inner Milky Way, the amplitude $c_0$ and $f_\mathrm{survive}$ are fully degenerate, but we nevertheless marginalize over them. The resulting posterior, shown in Fig. \ref{fig:mwdm_PDF}, sets a lower bound of $m_{\rm{WDM}} > 3.6\,\mathrm{keV}$ at 95\% confidence. The PDF plateaus for $m_{\rm{WDM}} \gtrsim 10\,\mathrm{keV}$ indicating masses above that are equally preferred by the GD-1 and Pal 5 stream. The constraint of $m_{\rm{WDM}} > 3.6\,\mathrm{keV}$ is the same as the one we obtained in B21 from fully forward modeling of the stream data using the WDM mass function, demonstrating that using the sufficient statistics of the streams SHMF measurements as we do in this paper captures all of the relevant information about the streams measurement. We do note that we found in B21 that the Pal 5 stream has a very weak preference for small WDM masses, weakening the combined constraint compared to one that only uses the GD-1 data. Using GD-1 alone, we found in B21 that $m_{\rm{WDM}} > 4.6\,\mathrm{keV}$ at 95\% confidence. 

\subsection{Constraints from stellar streams and dwarf galaxy counts}\label{sec:wdm_wdwarfs}

\begin{figure}
\centering
\includegraphics[width=0.75\textwidth]{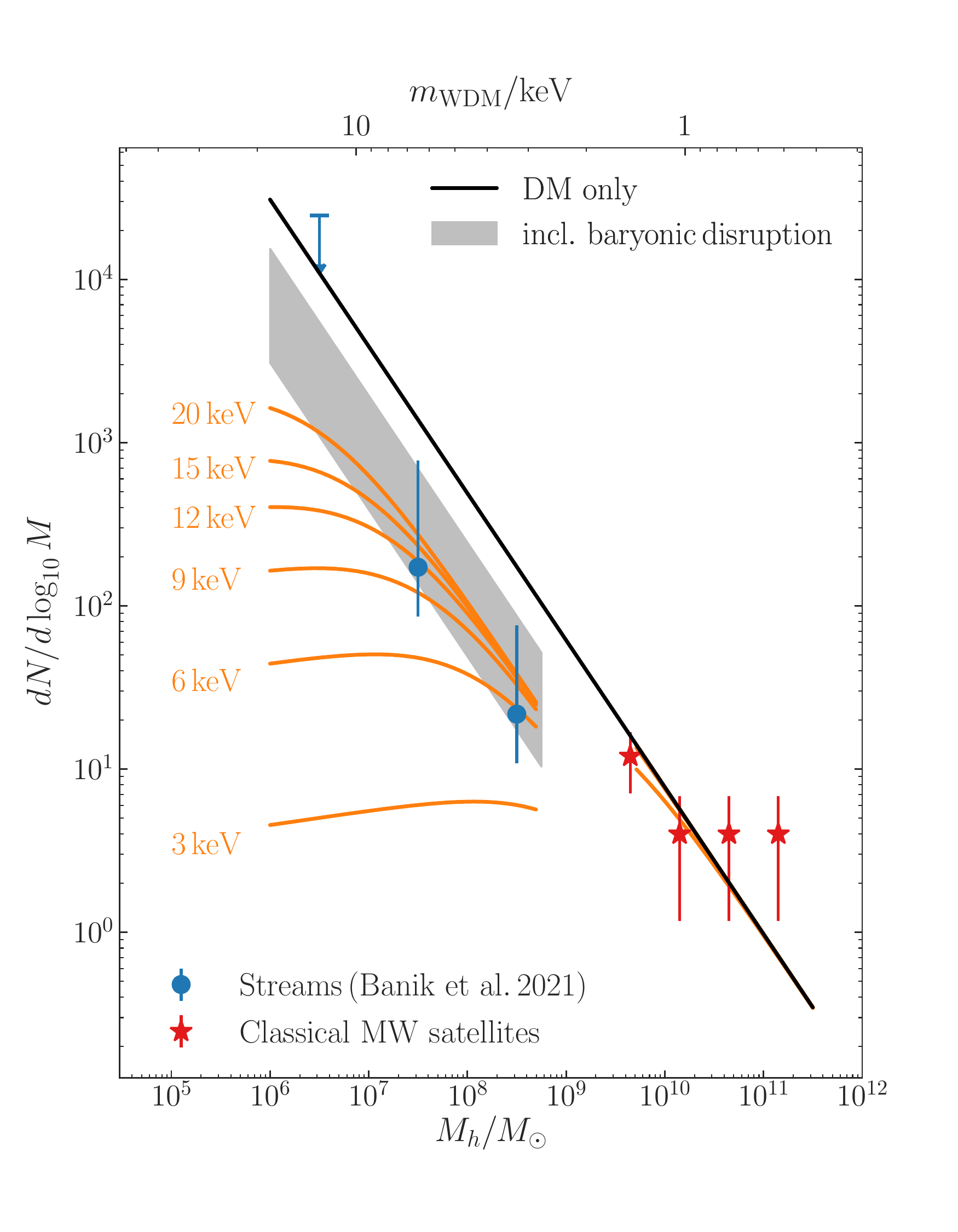}
\caption{SHMF in the mass range $10^{6} - 10^{9} \Msun$ reconstructed from the analysis of the perturbations induced on the GD-1 and Pal 5 streams. Red data points show the observed classical Milky Way satellites out to 300 kpc. The blue downward arrow and data points show the 68\% upper bound, and the measurement and 68\% error, respectively, in 3 mass bins below the scale of dwarfs, as obtained in B21 and extrapolated out to 300 kpc to place them on the same SHMF as the red points. The shaded area show the CDM mass function taking into account the baryonic disruption of the subhalos. The orange lines show the predicted mass function for thermal WDM candidates of different mass, taking into account the expected subhalo depletion due to baryonic disruption for the low-mass ($M < 10^9\Msun$) measurements from the inner Milky Way.}
\label{fig:subhalo_mass_function}
\end{figure}

\begin{figure}
\centering
\includegraphics[width=0.89\textwidth]{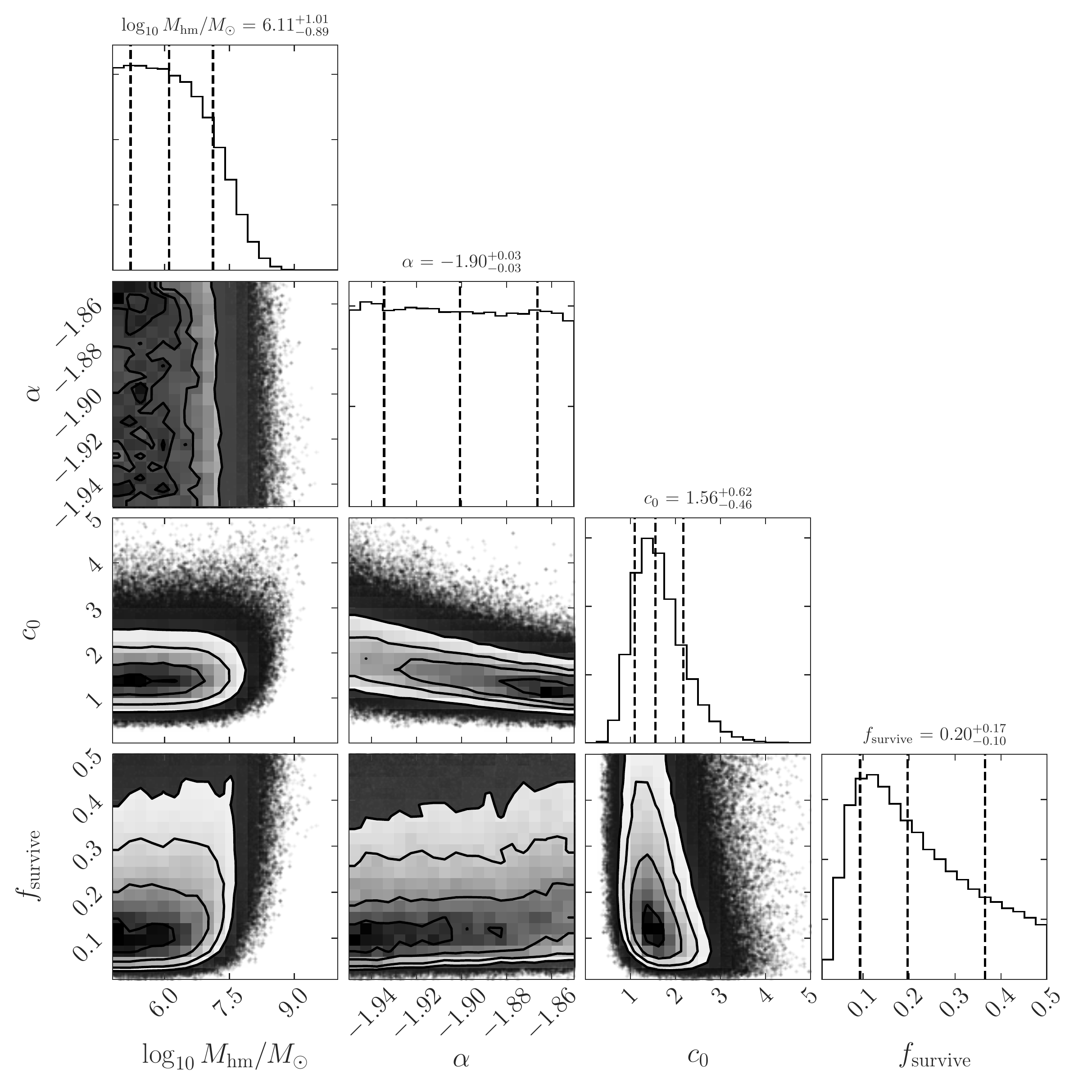}
\caption{Posterior PDFs for the WDM SHMF parameters when using constraints on the Milky Way's SHMF from both streams and classical-satellite counts. Made using \texttt{corner} \cite{corner}.}
\label{fig:full_PDF}
\end{figure}

In B21, we obtained a measurement of the SHMF in the mass range $10^{6} - 10^{9} \Msun$ using a joint analysis of the GD-1 and Pal 5 streams using the modeling described above. To extend the observed range of the SHMF and better constrain its amplitude in the unsuppressed regime, we add measurements from the counts of classical Milky Way satellites.  For these classical-satellite counts, we use the compilation of properties of the classical satellites (with stellar mass $> 10^5\,M_\odot$) within 300 kpc from Ref. \cite{GarrisonKimmel19a} and we assign dark-halo masses for all of these satellites using the stellar-mass vs. halo-mass relation
for satellites given in Ref. \cite{2017ARA&A..55..343B}. In Fig. \ref{fig:subhalo_mass_function}, we then display the Milky Way's observed SHMF of subhalos within 300 kpc, where we have extrapolated the SHMF at $M < 10^9\,M_\odot$ measured using streams within 23 kpc to 300 kpc using the assumed Einasto profile of the subhalo distribution from Section \ref{sec:subhalos}. However, because these $M < 10^9\,M_\odot$ SHMF measurements are actually made within 23 kpc, we expect them to be affected by baryonic tidal disruption and we account for this using the $f_\mathrm{survive}$ parameter in our analysis. The shaded region in Fig. \ref{fig:subhalo_mass_function} shows the range $f_\mathrm{survive} = [0.1,0.5]$. Note that we only perform the extrapolation discussed here for plotting purposes, in our actual fits we do not extrapolate. In this figure, we also use $c_0 = 1.6$, because we will find this to be the best fit below.

Fig. \ref{fig:subhalo_mass_function} also includes a number of representative thermal WDM SHMFs (orange lines). These take into account the expected subhalo depletion by $\approx80\,\%$ due to baryonic disruption for the low-mass, $M < 10^9\Msun$ measurements from the inner Milky Way (in practice, we use $f_\mathrm{survive} = 0.22$, the geometric mean of the bounds of the shaded region). It is clear that the SHMF measurements at low mass allows us to set a stringent lower limit on the particle mass $m_{\rm{WDM}}$.

\begin{table}
    \centering
    \begin{tabular}{|c|c|c|}
    \hline
        WDM mass function & Prior on $m_{\rm{WDM}}$ & 95\% lower bound on $m_{\rm{WDM}}$ \\
        \hline
        Lovell et al. 2014 \cite{Lovell2013} &  flat in $1/m_{\rm{WDM}}$ & 4.9 keV \\
        Lovell et al. 2014 \cite{Lovell2013} &  flat in $\log M_\mathrm{hm}$ & 6.2 keV \\
        \hline
        Lovell et al. 2021 \cite{Lovell21a} & flat in $1/m_{\rm{WDM}}$ & 2.1 keV \\
        Lovell et al. 2021 \cite{Lovell21a} &  flat in $\log M_\mathrm{hm}$ & 3.8 keV \\
        \hline
    \end{tabular}
    \caption{Constraint on the mass of a thermal WDM candidate arising from the comparison of the predicted mass function with the one inferred from the analysis of GD-1 and Pal 5 perturbations and from the counts of classical Milky Way satellites. The different rows show combinations of different priors on the WDM mass and different assumed forms of the WDM SHMF.}
    \label{tab:mwdm}
\end{table}

We fit our full model using a prior that is flat in $\log M_\mathrm{hm}$ as in the previous subsection and obtain the posterior distribution shown in Fig. \ref{fig:full_PDF}. The amplitude of the SHMF, $c_0$, is quite well constrained and the survival fraction $f_\mathrm{survive}$ is as well, with a posterior mean that is almost exactly at the geometric mean of the bounds of the shaded region in Fig. \ref{fig:subhalo_mass_function}. The logarithmic slope of the SHMF is only poorly constrained within its narrow prior range. Lastly, we clearly rule out high values of $M_\mathrm{hm}$ and obtain a stringent upper limit. Translated to a lower limit on $m_\mathrm{WDM}$ (PDF shown in Fig. \ref{fig:mwdm_PDF}), we find that $m_\mathrm{WDM} > 6.2\,\mathrm{keV}$ at 95\% confidence. The disadvantage of the prior that is flat in $\log M_\mathrm{hm}$ is that it is improper without a cut-off at low masses (which we placed at $\log_{10} M_\mathrm{hm} = 4.8$). To avoid this, we can also run with a prior that is uniform in $1/m_\mathrm{WDM}$. In this case, we find that $m_\mathrm{WDM} > 4.9\,\mathrm{keV}$. It is unsurprising that the limit in this case is less strong, because there is more prior weight given to regions of low WDM mass, so they are more difficult to rule out. Because the prior that is flat in $\log M_\mathrm{hm}$ is the standard assumption in the literaure (e.g., Ref. \cite{Gilman:2019nap}), we consider this our primary result.

As discussed in Section~\ref{sec:wdm}, recent WDM simulation work has indicated that the WDM mass function as probed by stellar streams may be less suppressed relative to CDM for a given WDM mass, because the inner regions of WDM halos may be primarily populated by tidally-stripped remnants of higher-mass subhalos (that are unsuppressed relative to CDM). As argued in Section \ref{sec:wdm}, we can approximately model this by changing the $\gamma$ and $\beta$ parameters in the WDM SHMF from Equation \eqref{eq:wdmcdm}. Running our fits using $\gamma = 1.5$ and $\beta = 0.7$, we find that $m_\mathrm{WDM} > 3.8\,\mathrm{keV}$ and $m_\mathrm{WDM} > 2.1\,\mathrm{keV}$ for priors that are flat in $\log M_\mathrm{hm}$ or $1/m_\mathrm{WDM}$, respectively. As expected, the less strong suppression of the WDM mass function leads to a less stringent lower bound on the WDM mass. However, for the standard prior that is flat in $\log M_\mathrm{hm}$, we still rule out the canonical $3.3\,\mathrm{keV}$ model at better than 95\% confidence. We summarize all of these constraints in Table \ref{tab:mwdm}.

\section{Constraints on fuzzy dark matter}

\begin{figure}
\centering
\includegraphics[width=0.8\textwidth]{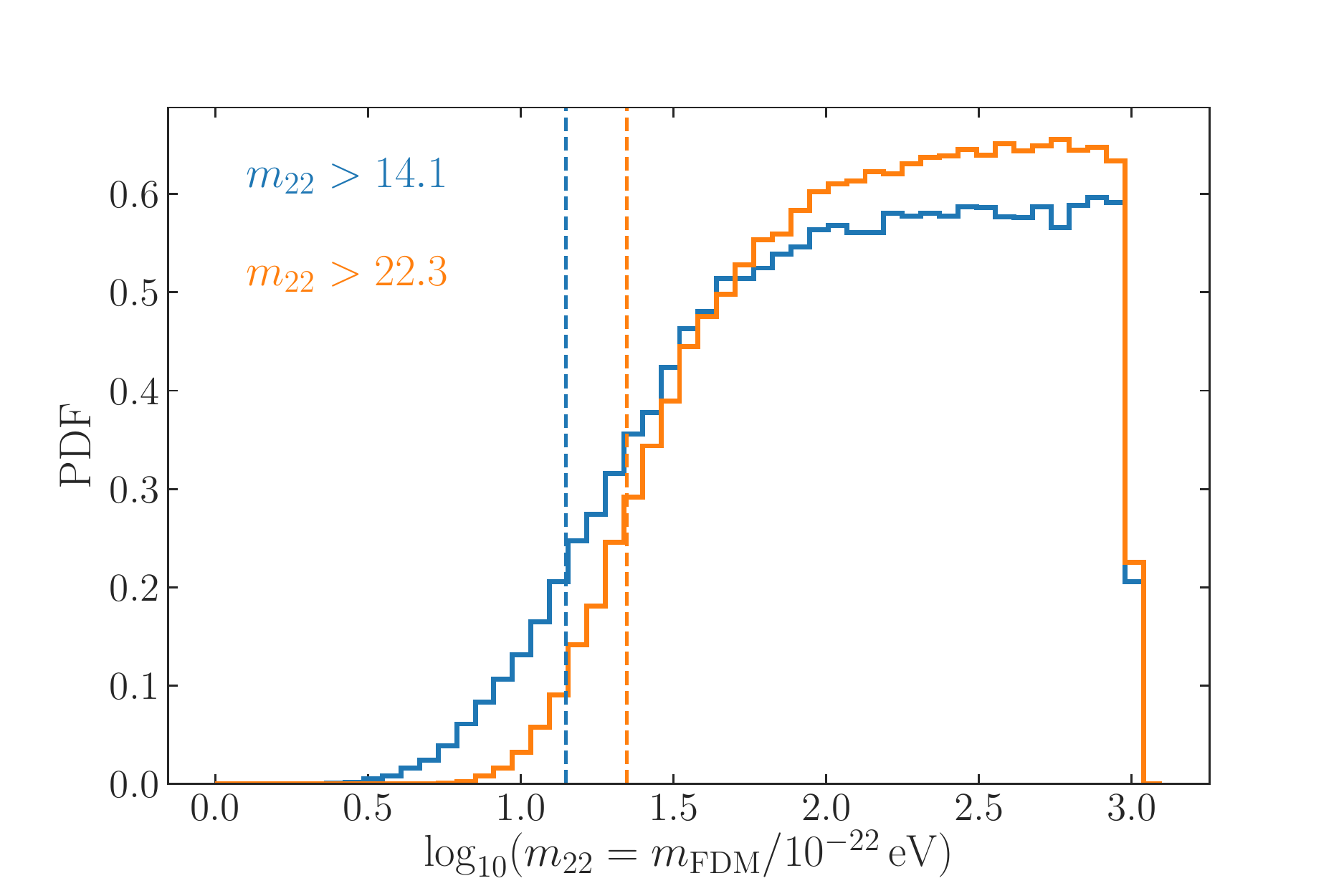}
\caption{Posterior PDF for the fuzzy DM particle mass using density fluctuations in the Pal 5 and GD-1 streams alone (blue curves) and combining this with classical-satellite counts (orange curves). The 95\% lower bounds of both PDFs are indicated with the labels and dashed vertical lines.}
\label{fig:mfdm_PDF}
\end{figure}

\begin{table}
    \centering
    \begin{tabular}{|c|c|}
    \hline
        Prior on $m_{\rm{FDM}}$ & 95\% lower bound on $m_{22} = m_{\rm{FDM}} / 10^{-22}\,\mathrm{eV}$ \\
        \hline
        flat in $1/m_{\rm{FDM}}$ & 12 \\
        flat in $\log m_{\rm{FDM}}$ & 22 \\
        \hline
    \end{tabular}
    \caption{Constraint on the FDM mass arising from the comparison of the predicted mass function with the one inferred from the analysis of GD-1 and Pal 5 perturbations and from the counts of classical Milky Way satellites. The different rows show different priors on the FDM mass.}
    \label{tab:mfdm}
\end{table}

\begin{figure}
\centering
\includegraphics[width=0.75\textwidth]{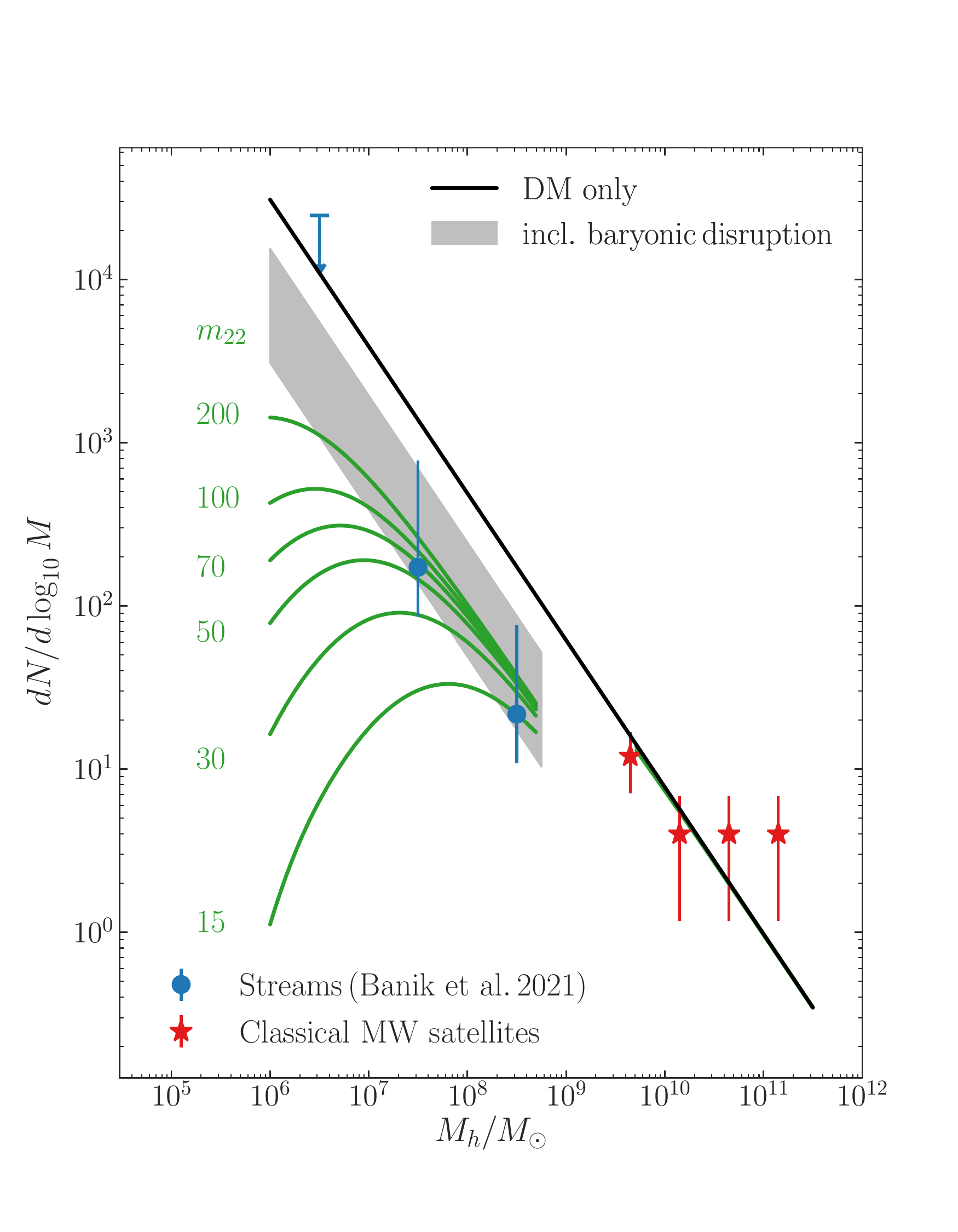}
\caption{Milky Way SHMF compared with fuzzy DM models for different FDM masses. Data, black line, and gray band are as in Fig. \ref{fig:subhalo_mass_function}, but green curves now show predicted SHMFs for fuzzy DM models with different FDM masses $m_{22} = m_\mathrm{FDM} / 10^{-22}\,\mathrm{eV}$.}
\label{fig:subhalo_mass_function_fdm}
\end{figure}

We now apply the same formalism that we applied to constrain the WDM mass to fuzzy DM.

\subsection{Constraints from streams alone}

As in Section \ref{sec:wdm_wstreams}, we start by considering the constraint on the FDM mass that we obtain from considering the Pal 5 and GD-1 stream data on its own. We again fit the four parameter model, which now uses the FDM mass in addition to the amplitude $c_0$ and logarithmic slope $\alpha$ of the base SHMF, and the factor $f_\mathrm{survive}$. Using a prior that is flat in $\log_{10} m_{22}$ between $m_{22} = 1$ and 1000, we obtain the PDF displayed in Fig. \ref{fig:mfdm_PDF}. The 95\% confidence lower bound on $m_{22}$ is 14.

\subsection{Constraints from stellar streams and dwarf galaxy counts}

\begin{figure}
\centering
\includegraphics[width=0.89\textwidth]{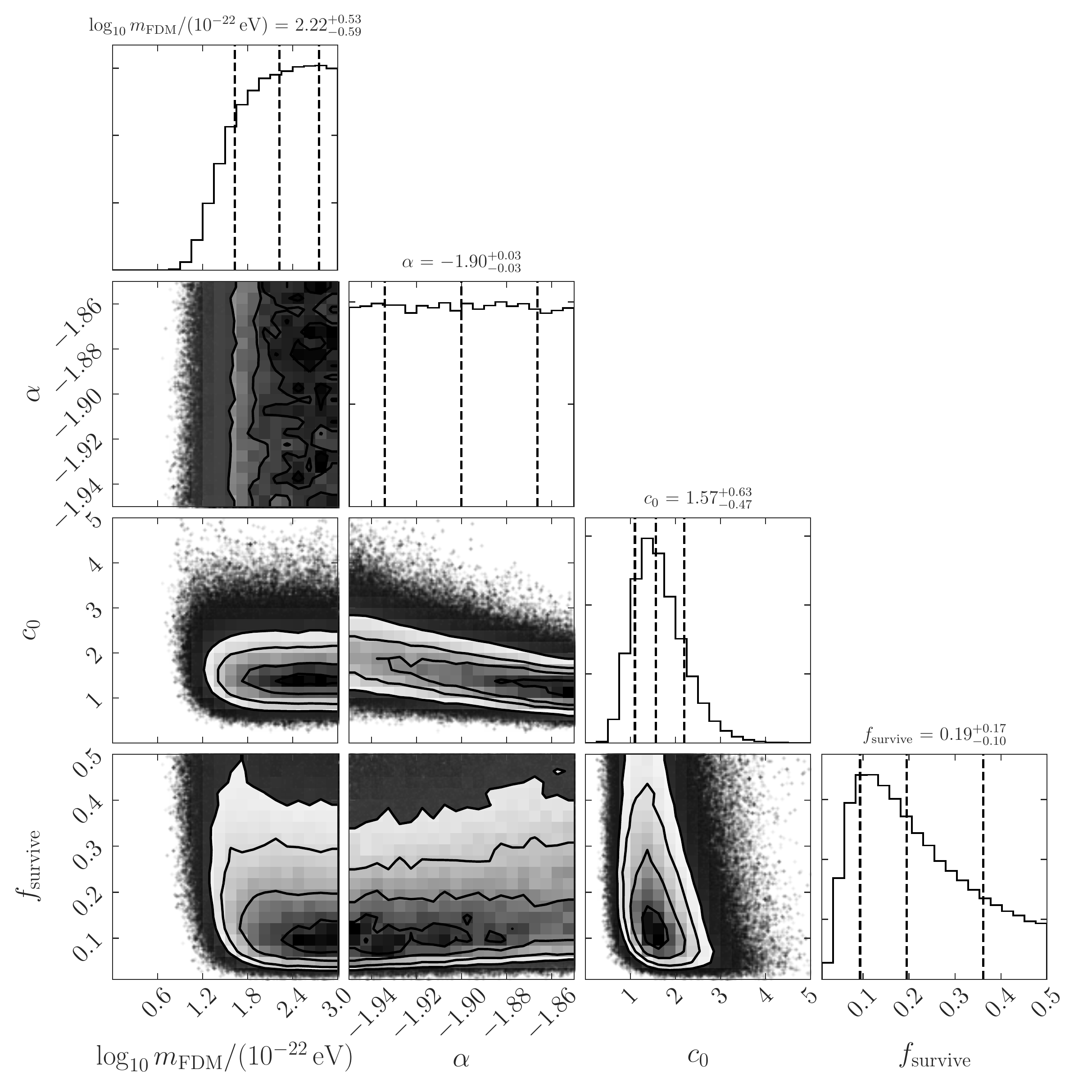}
\caption{Posterior PDFs for the FDM SHMF parameters when using constraints on the Milky Way's SHMF from both streams and classical-satellite counts. Made using \texttt{corner} \cite{corner}.}
\label{fig:full_PDF_fdm}
\end{figure}

To improve the constraint on the FDM mass coming from the observed SHMF, we again add measurements of the SHMF at higher masses using counts of classical satellites of the Milky Way in the same way as in Section \ref{sec:wdm_wdwarfs}. Fig. \ref{fig:subhalo_mass_function_fdm} compares the observed SHMF out to 300 kpc (obtained using extrapolation in the way described in Sec. \ref{sec:wdm_wdwarfs}) with FDM models with a range of values of $m_{22}$. Comparing the FDM SHMFs in this figure to the WDM SHMFs in Fig. \ref{fig:subhalo_mass_function}, we see that the suppression at low masses is much stronger in FDM. This means that the stream SHMF determinations on their own more strongly constrain $m_{22}$ than is the case for WDM, because there is a large difference in the predicted FDM SHMF between the two mass bins with stream measurements for relevant FDM models.

We fit the full model for the FDM SHMF to the stream and classical-dwarf data and the PDF for all of the parameters when using a prior that is flat in $\log m_{22}$ is shown in Fig. \ref{fig:full_PDF_fdm}. Comparing this figure to Fig. \ref{fig:full_PDF} for WDM, we see that the posterior distribution of the SHMF parameters $c_0$, $\alpha$, and $f_\mathrm{survive}$ is almost exactly the same as for WDM. This is unsurprising, given that good fits to the data are models that are close to CDM in both cases. We summarize the lower bound on $m_{22}$ at 95\% confidence resulting from this analysis in Table \ref{tab:mfdm}: using a prior that is flat in $\log m_{22}$, we find that $m_{22} > 22$, whereas with one that is flat in $1/m_{22}$, we find that $m_{22} > 12$. 

\section{Discussion and Conclusions} 

In B21, we found evidence for density perturbations on angular scales larger than $10^\circ$ in the observed linear density of the GD-1 stream, and we showed that a population of canonical CDM subhalos are necessary and sufficient to explain those perturbations and similar perturbations in the Pal 5 stream. Here, we have studied the impact of these findings on two popular alternative DM models: a thermal relic warm DM model and ultra-light axion DM (``fuzzy DM''). To better constrain the amplitude of the SHMF where it is relatively unsuppressed compared to the standard CDM expectation, we added constraints from counts of classical satellites. To sum up, we compare the limits that we derived to those obtained using other methods and we conclude with a more general discussion of how SHMF measurements can be used to constrain a large variety of DM models.

WDM has long been a popular DM candidate and traditionally it has largely been constrained using observations of the fluctuations in the Lyman-$\alpha$ forest (e.g., Ref. \cite{Viel13a} finds $m_\mathrm{WDM} > 3.3\,\mathrm{keV}$ at 95\% confidence; all limits below are also at 95\% confidence). The most recent constraint coming from such an analysis finds that  $m_\mathrm{WDM} > 5.3\,\mathrm{keV}$ \cite{Irsic17a}. Lyman-$\alpha$ forest constraints on WDM are unlikely to improve significantly in the future, because (baryonic) pressure smoothing of the intergalactic medium means that WDM acts like CDM for higher WDM masses as far as the Lyman-$\alpha$ forest is concerned. Equally or more stringent constraints have therefore more recently been obtained using SHMF measurements in elliptical galaxies from flux-ratio anomalies in strong gravitational lenses (Ref. \cite{Gilman:2019nap}: $m_\mathrm{WDM} > 5.2\,\mathrm{keV}$; Ref. \cite{Hsueh20a}: $m_\mathrm{WDM} > 5.6\,\mathrm{keV}$) and detailed modeling of the satellite population in the Milky Way (Ref. \cite{Nadler21a}: $m_\mathrm{WDM} > 6.5\,\mathrm{keV}$). A combined analysis of lensing and satellite constraints finds the most stringent constraint yet: $m_\mathrm{WDM} > 9.7\,\mathrm{keV}$ \cite{Nadler21b}; another joint analysis that also include Ly-$\alpha$ forest constraints finds a weaker constraint: $m_\mathrm{WDM} > 6\,\mathrm{keV}$ \cite{Enzi21a}. With the fiducial WDM mass function from Ref.~\cite{Lovell2013}, our constraint is $m_\mathrm{WDM} > 6.2\,\mathrm{keV}$ when using the same or a similar prior as the other SHMF constraints and the constraint from stellar streams is therefore on par with those of other SHMF measurements. However, if the abundance of WDM subhalos is less suppressed in the inner Milky Way because of tidal stripping as recently claimed by Ref. \cite{Lovell21a}, our constraint weakens to $m_\mathrm{WDM} > 3.8\,\mathrm{keV}$. While the strong-lensing constraint is largely unaffected by the tidal-stripping effect, the satellite constraint from Ref. \cite{Nadler21a} likely \emph{is} weakened similar to ours, although likely not as much.

Of the three SHMF methods to constrain WDM, stellar streams most directly measure the masses of three-dimensional DM subhalos. Using the abundance of Milky-Way satellites below the classical-satellite regime requires intricate modeling of the relation between stellar and DM halo mass, while strong lensing flux-ratio anomalies only measure projected mass and only in the aggregate. As such, lensing has a hard time distinguishing between a less abundant population of more concentrated subhalos and a more abundant population with lower concentration \cite{Gilman20b}. Perturbations to stellar streams, on the other hand, are not that sensitive to the internal structure of the DM subhalos \cite{Bovy2016a} and are mainly sensitive to their masses. Going forward, both the stellar streams and strong lensing measurements will be able to directly constrain lower subhalo masses by at least an order of magnitude, through using more streams and more detailed maps of them in the case of streams and by using more systems, mid-infrared flux ratios \cite{Gilman21a}, and detailed imaging \cite{Vegetti10a} for strong lenses. 

\begin{figure}
\centering
\includegraphics[width=0.85\textwidth]{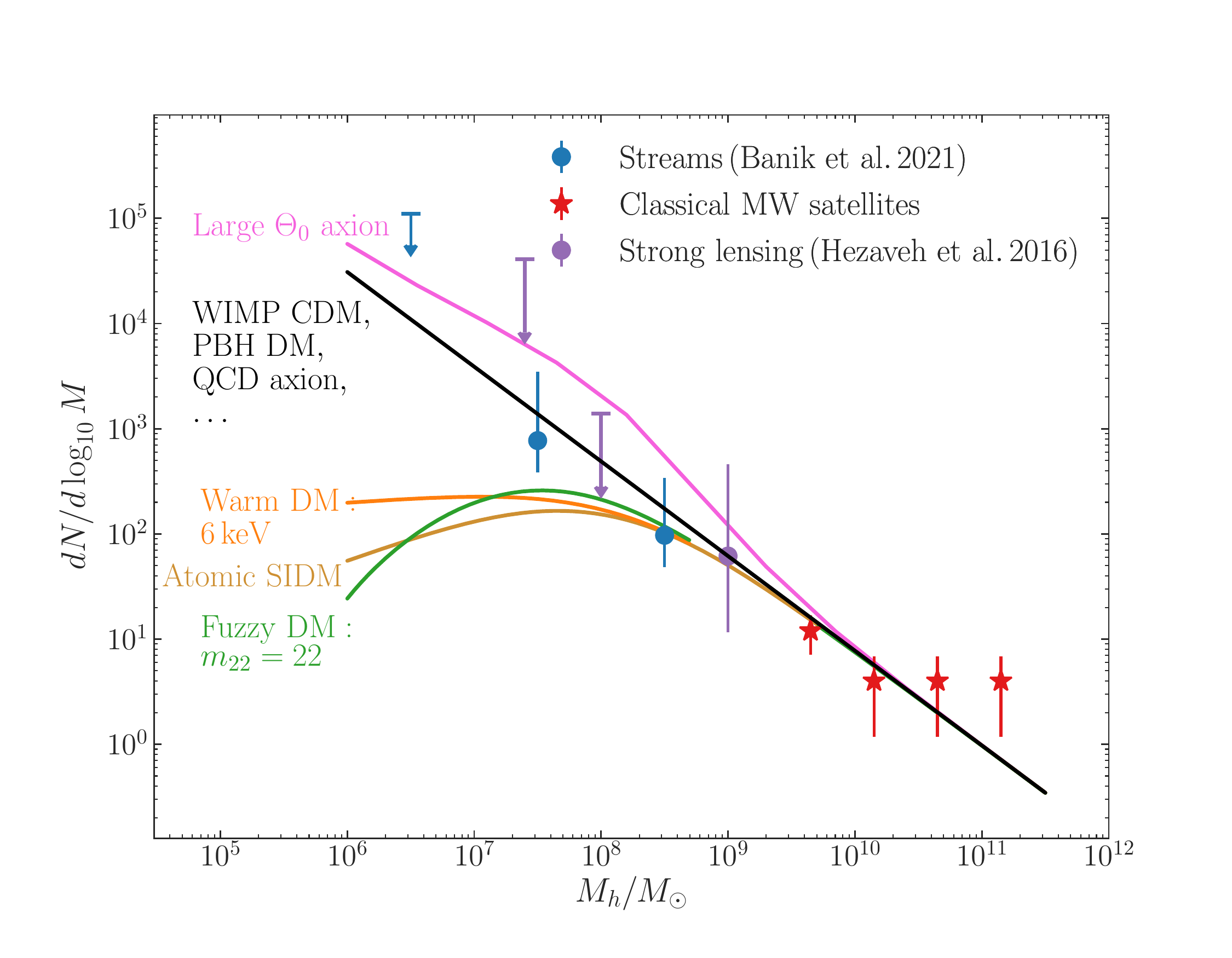}
\caption{SHMF of the Milky Way out to 300 kpc compared to different alternative DM models to illustrate the stringent bounds that the observed SHMF sets on deviations from the canonical CDM scenario. We include SHMF measurements from the GD-1 and Pal 5 streams, corrected for the effect of baryonic disruption; from counts of classical satellites in the Milky Way; and from strong-lensing measurements in external galaxies (these have been translated to Milky-Way SHMF constraints as discussed in the text). We include models that acts as the canonical CDM model: WIMP CDM, primordial black holes, and the QCD axion (black curve), as well as models that lead to a suppression of the SHMF on small scales (WDM and FDM at the limit derived in this paper and an ``atomic SIDM'' model that acts like WDM from Ref. \cite{Boddy16a}) or a higher abundance (the large-misalignment axion DM model from Ref. \cite{Arvanitaki:2019rax}). The observed SHMF strongly constrains models that predict different abundances of low-mass DM subhalos than the canonical CDM model.}
\label{fig:subhalo_mass_function_altdm}
\end{figure}

Because systematics and degeneracies in the different SHMF-based methods are largely different, combinations of constraints will remain highly informative. For example, combining our own fiducial constraint with that of the joint lensing+satellites analysis of Ref. \cite{Nadler21b} gives $m_\mathrm{WDM} \gtrsim 11\,\mathrm{keV}$ simply from multiplying the marginalized PDFs for $M_\mathrm{hm}$ (for this, we approximate the PDF from Ref. \cite{Nadler21b} as $p(\log_{10} M_\mathrm{hm})\propto\exp\left[-(\log_{10} M_\mathrm{hm} - 4.5)^{2.35}/5.7\right]$ over the range $\log_{10} M_\mathrm{hm} \in [5,9]$). Because the satellites part of the constraint from Ref. \cite{Nadler21b} involves the same Milky Way SHMF as our constraint, a full joint analysis would likely lead to an even stronger constraint (as we use the classical satellites, there is of course some double-counting of the data, but because Ref. \cite{Nadler21b} largely depend on satellites with masses below those that we consider, this likely does not have a big effect).

FDM models with masses $\gtrsim 10^{-23}\,\mathrm{eV}$ have similarly been constrained using the SHMF and the Ly-$\alpha$ forest data. Ref. \cite{Schutz20a} translated the stream measurements from B21 and the strong-lensing constraints from Ref. \cite{Gilman:2019nap} to a constraint on the FDM mass, finding $m_{22} > 21$. This was similar to existing Ly-$\alpha$ forest constraints: $m_{22}\gtrsim 21$ \cite{Armengaud17a,Irsic17b,Kobayashi17a}. Ref. \cite{Amorisco18a} constrained $m_{22} > 1.5$ from the absence of heating due to the expected wave-interference density fluctuations in low-mass FDM models. Recently, Ref. \cite{Nadler21a} used constraints from the Milky Way satellite population to set a slightly more stringent lower bound of $m_{22} > 29$, but a recent re-analysis of the Ly-$\alpha$ forest data gives $m_{22} > 200$, a much tighter constraint than that coming from the SHMF. Our own constraint ($m_{22} > 22$) is similar to the existing SHMF constraints, but falls far short of the Ly-$\alpha$ constraint. In our FDM modeling, we have also only included DM substructure in the form of subhalos, but FDM halos contain a fluctuating pattern of density fluctuations due to wave interference, which itself may heat and perturb stellar streams \cite{Amorisco2016,Dalal21a} and affect strong-lensing flux-ratio anomalies \cite{Chan20a}. While this effect should be small at $m_{22} \gtrsim 10$, a full analysis is nevertheless necessary to fully trust the constraint arising from density perturbations of stellar streams.

Measurements of the SHMF in the Milky Way and external galaxies provide strong constraints on wide classes of DM models beyond those that we have considered in this paper. To illustrate the power of SHMF measurements from streams, dwarf satellites, and strong lensing, we present in Fig. \ref{fig:subhalo_mass_function_altdm} a version of Figs. \ref{fig:subhalo_mass_function} and \ref{fig:subhalo_mass_function_fdm} where we have included constraints from strong lensing \cite{Hezaveh2016}. To remove the effect of baryonic disruption on the streams-based SHMF, we have corrected the streams-based SHMF measurement by multiplying them by the inverse of $f_\mathrm{survive} = 0.22$. We have included the lensing measurements simply by interpreting their constraints on the abundance of DM subhalos relative to the CDM prediction (in their more massive host halos) as constraints on the abundance of DM subhalos in the Milky Way relative to the CDM expectation. We stress that this is \emph{not} the proper way of combining constraints from lensing in external galaxies with measurements of the Milky Way's SHMF, but we do this simply to illustrate the complementarity and consistency of the strong-lensing and Milky-Way measurements. We also indicate where a variety of DM models lie in this space. Models that act like the canonical CDM model on the mass scales probed by the SHMF include the standard WIMP model \cite{Jungman:1995df,Bertone05}, primordial black holes \cite{Carr16a}, and the QCD axion \cite{Preskill83a,Abbott83a,Dine83a}. Models that predict a suppression of the SHMF at small scales include the thermal-relic WDM and FDM models that we considered in this paper, but also non-thermal WDM such as sterile neutrinos \cite{Dodelson1994} or certain self-interacting DM (SIDM) models with DM that acts like hydrogen \cite{Boddy16a,Huo18a} and are effectively warm (``Atomic SIDM'' in the figure, a model with $m_\mathrm{DM} = 50\,\mathrm{GeV}$, a dark electron 160 times lighter, and dark fine structure constant of $1/40$ \cite{Boddy16a}). Finally, some DM models predict more structure on small scales than that predicted by the canonical CDM model; as an example of this class of models we include axion DM produced through the large-misalignment mechanism \cite{Arvanitaki:2019rax} (the predicted SHMF in this scenario is an estimate using the transfer function and extended Press-Schechter calculations; Dalal \& Van Tilburg, private communication). 

It is clear that the SHMF measurements considered in this paper and those arising from strong lensing strongly constrain a wide range of non-CDM models. To encourage particle physicists working on alternative DM models to take into account constraints from the SHMF on their models, we release a \texttt{jupyter} notebook that reproduces all of the analysis in this paper and that is easily adapted to other SHMF models at: \\
\centerline{ \url{https://github.com/jobovy/dark-matter-constraints-from-stellar-streams}~,}\\
and further archived at\\
\centerline{ \url{https://zenodo.org/record/5234377\#.YSLwUlNKgb0}~.}

{\bf Acknowledgements.---} We thank Neal Dalal, Asimina Arvanitaki, and Ken Van Tilburg for providing us with the large-misalignment axion DM SHMF and Manoj Kaplinghat for providing us with the ``Atomic SIDM'' SHMF in Fig. \ref{fig:subhalo_mass_function_altdm}. We also thank Daniel Gilman and the anonymous referee for useful feedback. NB acknowledges the support of the D-ITP consortium, a programme of the Netherlands Organization for Scientific Research (NWO) that is funded by the Dutch Ministry of Education, Culture and Science (OCW). JB received support from the Natural Sciences and Engineering Research Council of Canada (NSERC; funding reference numbers RGPIN-2015-05235 and RGPIN-2020-04712) and an Ontario Early Researcher Award (ER16-12-061).

\bibliographystyle{JHEP}
\bibliography{refs}{}

\providecommand{\href}[2]{#2}\begingroup\raggedright\begin{thebibliography}{100}

\bibitem{Rubin80a}
V.~C. {Rubin}, J.~{Ford}, W.~K., and N.~{Thonnard}, {\it {Rotational properties
  of 21 SC galaxies with a large range of luminosities and radii, from NGC 4605
  (R=4kpc) to UGC 2885 (R=122kpc).}},  {\em \apj} {\bf 238} (June, 1980)
  471--487.

\bibitem{Davis85a}
M.~{Davis}, G.~{Efstathiou}, C.~S. {Frenk}, and S.~D.~M. {White}, {\it {The
  evolution of large-scale structure in a universe dominated by cold dark
  matter}},  {\em \apj} {\bf 292} (May, 1985) 371--394.

\bibitem{2015PhRvD..92l3516A}
{\'E}.~{Aubourg}, S.~{Bailey}, J.~E. {Bautista}, F.~{Beutler}, V.~{Bhardwaj},
  D.~{Bizyaev}, M.~{Blanton}, M.~{Blomqvist}, A.~S. {Bolton}, J.~{Bovy},
  H.~{Brewington}, J.~{Brinkmann}, J.~R. {Brownstein}, A.~{Burden}, N.~G.
  {Busca}, W.~{Carithers}, C.-H. {Chuang}, J.~{Comparat}, R.~A.~C. {Croft},
  A.~J. {Cuesta}, K.~S. {Dawson}, T.~{Delubac}, D.~J. {Eisenstein},
  A.~{Font-Ribera}, J.~{Ge}, J.~M. {Le Goff}, S.~G.~A. {Gontcho}, J.~R. {Gott},
  J.~E. {Gunn}, H.~{Guo}, J.~{Guy}, J.-C. {Hamilton}, S.~{Ho}, K.~{Honscheid},
  C.~{Howlett}, D.~{Kirkby}, F.~S. {Kitaura}, J.-P. {Kneib}, K.-G. {Lee},
  D.~{Long}, R.~H. {Lupton}, M.~V. {Maga{\~n}a}, V.~{Malanushenko},
  E.~{Malanushenko}, M.~{Manera}, C.~{Maraston}, D.~{Margala}, C.~K. {McBride},
  J.~{Miralda-Escud{\'e}}, A.~D. {Myers}, R.~C. {Nichol}, P.~{Noterdaeme},
  S.~E. {Nuza}, M.~D. {Olmstead}, D.~{Oravetz}, I.~{P{\^a}ris},
  N.~{Padmanabhan}, N.~{Palanque-Delabrouille}, K.~{Pan},
  M.~{Pellejero-Ibanez}, W.~J. {Percival}, P.~{Petitjean}, M.~M. {Pieri},
  F.~{Prada}, B.~{Reid}, J.~{Rich}, N.~A. {Roe}, A.~J. {Ross}, N.~P. {Ross},
  G.~{Rossi}, J.~A. {Rubi{\~n}o-Mart{\'\i}n}, A.~G. {S{\'a}nchez},
  L.~{Samushia}, R.~T. {G{\'e}nova-Santos}, C.~G. {Sc{\'o}ccola}, D.~J.
  {Schlegel}, D.~P. {Schneider}, H.-J. {Seo}, E.~{Sheldon}, A.~{Simmons}, R.~A.
  {Skibba}, A.~{Slosar}, M.~A. {Strauss}, D.~{Thomas}, J.~L. {Tinker},
  R.~{Tojeiro}, J.~A. {Vazquez}, M.~{Viel}, D.~A. {Wake}, B.~A. {Weaver}, D.~H.
  {Weinberg}, W.~M. {Wood-Vasey}, C.~{Y{\`e}che}, I.~{Zehavi}, G.-B. {Zhao},
  and {BOSS Collaboration}, {\it {Cosmological implications of baryon acoustic
  oscillation measurements}},  {\em \prd} {\bf 92} (Dec., 2015) 123516,
  [\href{http://arxiv.org/abs/1411.1074}{{\tt arXiv:1411.1074}}].

\bibitem{Kids2017a}
F.~{K{\"o}hlinger}, M.~{Viola}, B.~{Joachimi}, H.~{Hoekstra}, E.~{van Uitert},
  H.~{Hildebrandt}, A.~{Choi}, T.~{Erben}, C.~{Heymans}, S.~{Joudaki},
  D.~{Klaes}, K.~{Kuijken}, J.~{Merten}, L.~{Miller}, P.~{Schneider}, and E.~A.
  {Valentijn}, {\it {KiDS-450: the tomographic weak lensing power spectrum and
  constraints on cosmological parameters}},  {\em \mnras} {\bf 471} (Nov.,
  2017) 4412--4435, [\href{http://arxiv.org/abs/1706.02892}{{\tt
  arXiv:1706.02892}}].

\bibitem{Planck2020a}
{Planck Collaboration}, N.~{Aghanim}, Y.~{Akrami}, M.~{Ashdown}, J.~{Aumont},
  C.~{Baccigalupi}, M.~{Ballardini}, A.~J. {Banday}, R.~B. {Barreiro},
  N.~{Bartolo}, S.~{Basak}, R.~{Battye}, K.~{Benabed}, J.~P. {Bernard},
  M.~{Bersanelli}, P.~{Bielewicz}, J.~J. {Bock}, J.~R. {Bond}, J.~{Borrill},
  F.~R. {Bouchet}, F.~{Boulanger}, M.~{Bucher}, C.~{Burigana}, R.~C. {Butler},
  E.~{Calabrese}, J.~F. {Cardoso}, J.~{Carron}, A.~{Challinor}, H.~C. {Chiang},
  J.~{Chluba}, L.~P.~L. {Colombo}, C.~{Combet}, D.~{Contreras}, B.~P. {Crill},
  F.~{Cuttaia}, P.~{de Bernardis}, G.~{de Zotti}, J.~{Delabrouille}, J.~M.
  {Delouis}, E.~{Di Valentino}, J.~M. {Diego}, O.~{Dor{\'e}}, M.~{Douspis},
  A.~{Ducout}, X.~{Dupac}, S.~{Dusini}, G.~{Efstathiou}, F.~{Elsner}, T.~A.
  {En{\ss}lin}, H.~K. {Eriksen}, Y.~{Fantaye}, M.~{Farhang}, J.~{Fergusson},
  R.~{Fernandez-Cobos}, F.~{Finelli}, F.~{Forastieri}, M.~{Frailis}, A.~A.
  {Fraisse}, E.~{Franceschi}, A.~{Frolov}, S.~{Galeotta}, S.~{Galli},
  K.~{Ganga}, R.~T. {G{\'e}nova-Santos}, M.~{Gerbino}, T.~{Ghosh},
  J.~{Gonz{\'a}lez-Nuevo}, K.~M. {G{\'o}rski}, S.~{Gratton}, A.~{Gruppuso},
  J.~E. {Gudmundsson}, J.~{Hamann}, W.~{Handley}, F.~K. {Hansen}, D.~{Herranz},
  S.~R. {Hildebrandt}, E.~{Hivon}, Z.~{Huang}, A.~H. {Jaffe}, W.~C. {Jones},
  A.~{Karakci}, E.~{Keih{\"a}nen}, R.~{Keskitalo}, K.~{Kiiveri}, J.~{Kim},
  T.~S. {Kisner}, L.~{Knox}, N.~{Krachmalnicoff}, M.~{Kunz}, H.~{Kurki-Suonio},
  G.~{Lagache}, J.~M. {Lamarre}, A.~{Lasenby}, M.~{Lattanzi}, C.~R. {Lawrence},
  M.~{Le Jeune}, P.~{Lemos}, J.~{Lesgourgues}, F.~{Levrier}, A.~{Lewis},
  M.~{Liguori}, P.~B. {Lilje}, M.~{Lilley}, V.~{Lindholm},
  M.~{L{\'o}pez-Caniego}, P.~M. {Lubin}, Y.~Z. {Ma}, J.~F.
  {Mac{\'\i}as-P{\'e}rez}, G.~{Maggio}, D.~{Maino}, N.~{Mandolesi},
  A.~{Mangilli}, A.~{Marcos-Caballero}, M.~{Maris}, P.~G. {Martin},
  M.~{Martinelli}, E.~{Mart{\'\i}nez-Gonz{\'a}lez}, S.~{Matarrese}, N.~{Mauri},
  J.~D. {McEwen}, P.~R. {Meinhold}, A.~{Melchiorri}, A.~{Mennella},
  M.~{Migliaccio}, M.~{Millea}, S.~{Mitra}, M.~A. {Miville-Desch{\^e}nes},
  D.~{Molinari}, L.~{Montier}, G.~{Morgante}, A.~{Moss}, P.~{Natoli}, H.~U.
  {N{\o}rgaard-Nielsen}, L.~{Pagano}, D.~{Paoletti}, B.~{Partridge},
  G.~{Patanchon}, H.~V. {Peiris}, F.~{Perrotta}, V.~{Pettorino},
  F.~{Piacentini}, L.~{Polastri}, G.~{Polenta}, J.~L. {Puget}, J.~P. {Rachen},
  M.~{Reinecke}, M.~{Remazeilles}, A.~{Renzi}, G.~{Rocha}, C.~{Rosset},
  G.~{Roudier}, J.~A. {Rubi{\~n}o-Mart{\'\i}n}, B.~{Ruiz-Granados},
  L.~{Salvati}, M.~{Sandri}, M.~{Savelainen}, D.~{Scott}, E.~P.~S. {Shellard},
  C.~{Sirignano}, G.~{Sirri}, L.~D. {Spencer}, R.~{Sunyaev}, A.~S. {Suur-Uski},
  J.~A. {Tauber}, D.~{Tavagnacco}, M.~{Tenti}, L.~{Toffolatti}, M.~{Tomasi},
  T.~{Trombetti}, L.~{Valenziano}, J.~{Valiviita}, B.~{Van Tent}, L.~{Vibert},
  P.~{Vielva}, F.~{Villa}, N.~{Vittorio}, B.~D. {Wandelt}, I.~K. {Wehus},
  M.~{White}, S.~D.~M. {White}, A.~{Zacchei}, and A.~{Zonca}, {\it {Planck 2018
  results. VI. Cosmological parameters}},  {\em \aap} {\bf 641} (Sept., 2020)
  A6, [\href{http://arxiv.org/abs/1807.06209}{{\tt arXiv:1807.06209}}].

\bibitem{Weinberg15a}
D.~H. {Weinberg}, J.~S. {Bullock}, F.~{Governato}, R.~{Kuzio de Naray}, and
  A.~H.~G. {Peter}, {\it {Cold dark matter: Controversies on small scales}},
  {\em Proceedings of the National Academy of Science} {\bf 112} (Oct., 2015)
  12249--12255, [\href{http://arxiv.org/abs/1306.0913}{{\tt arXiv:1306.0913}}].

\bibitem{2017ARA&A..55..343B}
J.~S. {Bullock} and M.~{Boylan-Kolchin}, {\it {Small-Scale Challenges to the
  {\ensuremath{\Lambda}}CDM Paradigm}},  {\em \araa} {\bf 55} (Aug, 2017)
  343--387, [\href{http://arxiv.org/abs/1707.04256}{{\tt arXiv:1707.04256}}].

\bibitem{Bond91a}
J.~R. {Bond}, S.~{Cole}, G.~{Efstathiou}, and N.~{Kaiser}, {\it {Excursion Set
  Mass Functions for Hierarchical Gaussian Fluctuations}},  {\em \apj} {\bf
  379} (Oct., 1991) 440.

\bibitem{Klypin99a}
A.~{Klypin}, A.~V. {Kravtsov}, O.~{Valenzuela}, and F.~{Prada}, {\it {Where Are
  the Missing Galactic Satellites?}},  {\em \apj} {\bf 522} (Sept., 1999)
  82--92, [\href{http://arxiv.org/abs/astro-ph/9901240}{{\tt
  astro-ph/9901240}}].

\bibitem{Moore99a}
B.~{Moore}, S.~{Ghigna}, F.~{Governato}, G.~{Lake}, T.~{Quinn}, J.~{Stadel},
  and P.~{Tozzi}, {\it {Dark Matter Substructure within Galactic Halos}},  {\em
  \apjl} {\bf 524} (Oct., 1999) L19--L22,
  [\href{http://arxiv.org/abs/astro-ph/9907411}{{\tt astro-ph/9907411}}].

\bibitem{Dodelson94a}
S.~{Dodelson} and L.~M. {Widrow}, {\it {Sterile neutrinos as dark matter}},
  {\em \prl} {\bf 72} (Jan., 1994) 17--20,
  [\href{http://arxiv.org/abs/hep-ph/9303287}{{\tt hep-ph/9303287}}].

\bibitem{Bode01a}
P.~{Bode}, J.~P. {Ostriker}, and N.~{Turok}, {\it {Halo Formation in Warm Dark
  Matter Models}},  {\em \apj} {\bf 556} (July, 2001) 93--107,
  [\href{http://arxiv.org/abs/astro-ph/0010389}{{\tt astro-ph/0010389}}].

\bibitem{Hui17a}
L.~{Hui}, J.~P. {Ostriker}, S.~{Tremaine}, and E.~{Witten}, {\it {Ultralight
  scalars as cosmological dark matter}},  {\em \prd} {\bf 95} (Feb., 2017)
  043541, [\href{http://arxiv.org/abs/1610.08297}{{\tt arXiv:1610.08297}}].

\bibitem{Nadler19a}
E.~O. {Nadler}, V.~{Gluscevic}, K.~K. {Boddy}, and R.~H. {Wechsler}, {\it
  {Constraints on Dark Matter Microphysics from the Milky Way Satellite
  Population}},  {\em \apjl} {\bf 878} (June, 2019) L32,
  [\href{http://arxiv.org/abs/1904.10000}{{\tt arXiv:1904.10000}}].

\bibitem{Gilman:2019nap}
D.~{Gilman}, S.~{Birrer}, A.~{Nierenberg}, T.~{Treu}, X.~{Du}, and A.~{Benson},
  {\it {Warm dark matter chills out: constraints on the halo mass function and
  the free-streaming length of dark matter with 8 quadruple-image strong
  gravitational lenses}},  {\em \apj} {\bf 897} (Aug, 2019) 147,
  [\href{http://arxiv.org/abs/1908.06983}{{\tt arXiv:1908.06983}}].

\bibitem{Ibata2001}
R.~A. Ibata, G.~F. Lewis, and M.~J. Irwin, {\it Uncovering cdm halo
  substructure with tidal streams},  {\em \mnras} {\bf 332} (June, 2002)
  915--920,
  [\href{http://arxiv.org/abs/http://arxiv.org/abs/astro-ph/0110690v1}{{\tt
  http://arxiv.org/abs/astro-ph/0110690v1}}].

\bibitem{Johnston2002}
K.~V. Johnston, D.~N. Spergel, and C.~Haydn, {\it How lumpy is the milky
  way’s dark matter halo?},  {\em \apj} {\bf 570} (may, 2002) 656.

\bibitem{Johnston1998}
K.~V. {Johnston}, {\it {A Prescription for Building the Milky Way's Halo from
  Disrupted Satellites}},  {\em \apj} {\bf 495} (Mar., 1998) 297--308,
  [\href{http://arxiv.org/abs/astro-ph/9710007}{{\tt astro-ph/9710007}}].

\bibitem{Eyre2011}
A.~Eyre and J.~Binney, {\it The mechanics of tidal streams},  {\em \mnras} {\bf
  413} (May, 2011) 1852--1874, [\href{http://arxiv.org/abs/1011.3672}{{\tt
  arXiv:1011.3672}}].

\bibitem{Bovy2014}
J.~Bovy, {\it Dynamical modeling of tidal streams},  {\em \apj} {\bf 795} (oct,
  2014) 95, [\href{http://arxiv.org/abs/1401.2985}{{\tt arXiv:1401.2985}}].

\bibitem{Yoon2011}
J.~H. Yoon, K.~V. Johnston, and D.~W. Hogg, {\it Clumpy streams from clumpy
  halos: detecting missing satellites with cold stellar structures},  {\em
  \apj} {\bf 731} (mar, 2011) 58.

\bibitem{Carlberg2012}
R.~G. Carlberg, {\it Dark matter sub-halo counts via star stream crossings},
  {\em \apj} {\bf 748} (mar, 2012) 20.

\bibitem{Carlberg2013}
R.~Carlberg, {\it The dynamics of star stream gaps},  {\em \apj} {\bf 775}
  (sep, 2013) 90.

\bibitem{Erkal2015}
D.~Erkal and V.~Belokurov, {\it Forensics of subhalo--stream encounters: the
  three phases of gap growth},  {\em \mnras} {\bf 450} (apr, 2015) 1136--1149.

\bibitem{Erkal2015a}
D.~Erkal and V.~Belokurov, {\it Properties of dark subhaloes from gaps in tidal
  streams},  {\em \mnras} {\bf 454} (oct, 2015) 3542--3558.

\bibitem{Sanders2016}
J.~L. Sanders, J.~Bovy, and D.~Erkal, {\it Dynamics of stream--subhalo
  interactions},  {\em \mnras} {\bf 457} (jan, 2016) 3817--3835.

\bibitem{Bovy2016a}
J.~Bovy, D.~Erkal, and J.~L. Sanders, {\it {Linear perturbation theory for
  tidal streams and the small-scale CDM power spectrum}},  {\em \mnras} {\bf
  466} (nov, 2017) 628--668, [\href{http://arxiv.org/abs/1606.03470}{{\tt
  arXiv:1606.03470}}].

\bibitem{Banik2020M}
N.~{Banik}, J.~{Bovy}, G.~{Bertone}, D.~{Erkal}, and T.~J.~L. {de Boer}, {\it
  {Evidence of a population of dark subhaloes from Gaia and Pan-STARRS
  observations of the GD-1 stream}},  {\em \mnras} {\bf 502} (Apr., 2021)
  2364--2380, [\href{http://arxiv.org/abs/1911.02662}{{\tt arXiv:1911.02662}}].

\bibitem{Banik2018}
N.~Banik, G.~Bertone, J.~Bovy, and N.~Bozorgnia, {\it Probing the nature of
  dark matter particles with stellar streams},  {\em \jcap} {\bf 7} (July,
  2018) 061,
  [\href{http://arxiv.org/abs/http://arxiv.org/abs/1804.04384v2}{{\tt
  http://arxiv.org/abs/1804.04384v2}}].

\bibitem{Dalal21a}
N.~{Dalal}, J.~{Bovy}, L.~{Hui}, and X.~{Li}, {\it {Don't cross the streams:
  caustics from fuzzy dark matter}},  {\em \jcap} {\bf 2021} (Mar., 2021) 076,
  [\href{http://arxiv.org/abs/2011.13141}{{\tt arXiv:2011.13141}}].

\bibitem{GAIAmain1}
{Gaia Collaboration}, A.~G.~A. {Brown}, and et~al., {\it {Gaia Data Release 1.
  Summary of the astrometric, photometric, and survey properties}},  {\em \aap}
  {\bf 595} (Nov., 2016) A2, [\href{http://arxiv.org/abs/1609.04172}{{\tt
  arXiv:1609.04172}}].

\bibitem{GAIAmain2}
{Gaia Collaboration}, T.~{Prusti}, and et~al., {\it {The Gaia mission}},  {\em
  \aap} {\bf 595} (Nov., 2016) A1, [\href{http://arxiv.org/abs/1609.04153}{{\tt
  arXiv:1609.04153}}].

\bibitem{Lindegren18}
L.~{Lindegren} and et~al., {\it {Gaia Data Release 2: The astrometric
  solution}},  {\em \aap} {\bf 616} (Apr., 2018) A2,
  [\href{http://arxiv.org/abs/1804.09366}{{\tt arXiv:1804.09366}}].

\bibitem{Chambers16}
K.~C. {Chambers} and et~al., {\it {The Pan-STARRS1 Surveys}},  {\em ArXiv
  e-prints} (Dec., 2016) [\href{http://arxiv.org/abs/1612.05560}{{\tt
  arXiv:1612.05560}}].

\bibitem{deboer_gd1_2018}
T.~J.~L. {de Boer}, V.~{Belokurov}, S.~E. {Koposov}, L.~{Ferrarese},
  D.~{Erkal}, P.~{C{\^o}t{\'e}}, and J.~F. {Navarro}, {\it {A deeper look at
  the GD1 stream: density variations and wiggles}},  {\em \mnras} {\bf 477}
  (Jun, 2018) 1893--1902, [\href{http://arxiv.org/abs/1801.08948}{{\tt
  arXiv:1801.08948}}].

\bibitem{Ibata2016}
R.~A. Ibata, G.~F. Lewis, and N.~F. Martin, {\it Feeling the pull, a study of
  natural galactic accelerometers. i: photometry of the delicate stellar stream
  of the palomar 5 globular cluster},  {\em \apj} {\bf 819} (feb, 2016) 1,
  [\href{http://arxiv.org/abs/http://arxiv.org/abs/1512.03054v1}{{\tt
  http://arxiv.org/abs/1512.03054v1}}].

\bibitem{Banik2019}
N.~Banik and J.~Bovy, {\it Effects of baryonic and dark matter substructure on
  the pal 5 stream},  {\em \mnras} {\bf 484} (jan, 2019) 2009--2020,
  [\href{http://arxiv.org/abs/http://arxiv.org/abs/1809.09640v2}{{\tt
  http://arxiv.org/abs/1809.09640v2}}].

\bibitem{Grillmair2006}
C.~J. Grillmair and O.~Dionatos, {\it Detection of a 63 deg cold stellar stream
  in the sloan digital sky survey},  {\em \apj} {\bf 643} (may, 2006) L17--L20,
  [\href{http://arxiv.org/abs/astro-ph/0604332}{{\tt astro-ph/0604332}}].

\bibitem{Price-Whelan18a}
A.~M. {Price-Whelan} and A.~{Bonaca}, {\it {Off the Beaten Path: Gaia Reveals
  GD-1 Stars outside of the Main Stream}},  {\em \apj} {\bf 863} (Aug., 2018)
  L20, [\href{http://arxiv.org/abs/1805.00425}{{\tt arXiv:1805.00425}}].

\bibitem{Webb2018}
J.~J. Webb and J.~Bovy, {\it Searching for the gd-1 stream progenitor in gaia
  dr2 with direct n-body simulations},  {\em arXiv e-prints} (Nov., 2018)
  [\href{http://arxiv.org/abs/http://arxiv.org/abs/1811.07022v1}{{\tt
  http://arxiv.org/abs/1811.07022v1}}].

\bibitem{deBoer20a}
T.~J.~L. {de Boer}, D.~{Erkal}, and M.~{Gieles}, {\it {A closer look at the
  spur, blob, wiggle, and gaps in GD-1}},  {\em \mnras} {\bf 494} (Apr., 2020)
  5315--5332, [\href{http://arxiv.org/abs/1911.05745}{{\tt arXiv:1911.05745}}].

\bibitem{Bovy2015}
J.~Bovy, {\it galpy: A python library for galactic dynamics},  {\em \apjs} {\bf
  216} (Feb., 2015) 29, [\href{http://arxiv.org/abs/1412.3451}{{\tt
  arXiv:1412.3451}}].

\bibitem{Kuepper2009}
A.~H.~W. Kuepper, P.~Kroupa, H.~Baumgardt, and D.~C. Heggie, {\it Tidal tails
  of star clusters},  {\em \mnras} {\bf 401} (jan, 2010) 105--120,
  [\href{http://arxiv.org/abs/http://arxiv.org/abs/0909.2619v2}{{\tt
  http://arxiv.org/abs/0909.2619v2}}].

\bibitem{Kuepper2011}
A.~H.~W. Kuepper, R.~R. Lane, and D.~C. Heggie, {\it More on the structure of
  tidal tails},  {\em \mnras} {\bf 420} (jan, 2012) 2700--2714,
  [\href{http://arxiv.org/abs/http://arxiv.org/abs/1111.5013v2}{{\tt
  http://arxiv.org/abs/1111.5013v2}}].

\bibitem{Ngan2013}
W.-H.~W. Ngan and R.~G. Carlberg, {\it Using gaps in n-body tidal streams to
  probe missing satellites},  {\em \apj} {\bf 788} (jun, 2014) 181,
  [\href{http://arxiv.org/abs/http://arxiv.org/abs/1311.1710v2}{{\tt
  http://arxiv.org/abs/1311.1710v2}}].

\bibitem{Koposov2010}
S.~E. Koposov, H.-W. Rix, and D.~W. Hogg, {\it Constraining the milky way
  potential with a six-dimensional phase-space map of the gd-1 stellar stream},
   {\em \apj} {\bf 712} (feb, 2010) 260--273,
  [\href{http://arxiv.org/abs/0907.1085}{{\tt arXiv:0907.1085}}].

\bibitem{Odenkirchen:2003ga}
{\bf SDSS} Collaboration, M.~Odenkirchen, E.~K. Grebel, W.~Dehnen, H.-W. Rix,
  B.~Yanny, H.~Newberg, C.~M. Rockosi, D.~Martinez-Delgado, J.~Brinkmann, and
  J.~R. Pier, {\it {The Extended tails of Palomar 5: A Ten degree arc of
  globular cluster tidal debris}},  {\em \aj} {\bf 126} (2003) 2385,
  [\href{http://arxiv.org/abs/astro-ph/0307446}{{\tt astro-ph/0307446}}].

\bibitem{Starkman2020}
N.~{Starkman}, J.~{Bovy}, and J.~J. {Webb}, {\it {An extended Pal 5 stream in
  Gaia DR2}},  {\em \mnras} {\bf 493} (Apr., 2020) 4978--4986,
  [\href{http://arxiv.org/abs/1909.03048}{{\tt arXiv:1909.03048}}].

\bibitem{Erkal2017}
D.~Erkal, S.~E. Koposov, and V.~Belokurov, {\it A sharper view of pal 5's
  tails: Discovery of stream perturbations with a novel non-parametric
  technique},  {\em \mnras} {\bf 470} (may, 2017) 60--84,
  [\href{http://arxiv.org/abs/1609.01282}{{\tt arXiv:1609.01282}}].

\bibitem{Pearson2017}
S.~Pearson, A.~M. Price-Whelan, and K.~V. Johnston, {\it Gaps and length
  asymmetry in the stellar stream palomar 5 as effects of galactic bar
  rotation},  {\em Nature Astronomy} {\bf 1} (Sept., 2017) 633--639,
  [\href{http://arxiv.org/abs/1703.04627}{{\tt arXiv:1703.04627}}].

\bibitem{Amorisco2016}
N.~C. Amorisco, F.~A. G\`{o}mez, S.~Vegetti, and S.~D.~M. White, {\it Gaps in
  globular cluster streams: giant molecular clouds can cause them too},  {\em
  \mnras} {\bf 463} (jul, 2016) L17--L21,
  [\href{http://arxiv.org/abs/1606.02715}{{\tt arXiv:1606.02715}}].

\bibitem{Erkal17a}
D.~{Erkal}, S.~E. {Koposov}, and V.~{Belokurov}, {\it {A sharper view of Pal
  5's tails: discovery of stream perturbations with a novel non-parametric
  technique}},  {\em \mnras} {\bf 470} (Sept., 2017) 60--84,
  [\href{http://arxiv.org/abs/1609.01282}{{\tt arXiv:1609.01282}}].

\bibitem{Wang2012}
Y.~Wang, H.~Zhao, S.~Mao, and R.~M. Rich, {\it A new model for the milky way
  bar},  {\em \mnras} {\bf 427} (nov, 2012) 1429--1440,
  [\href{http://arxiv.org/abs/http://arxiv.org/abs/1209.0963v1}{{\tt
  http://arxiv.org/abs/1209.0963v1}}].

\bibitem{Portail2016}
M.~Portail, O.~Gerhard, C.~Wegg, and M.~Ness, {\it Dynamical modelling of the
  galactic bulge and bar: the milky way's bar pattern speed, stellar, and dark
  matter mass distribution},  {\em \mnras} {\bf 465} (nov, 2016) 1621--1644,
  [\href{http://arxiv.org/abs/http://arxiv.org/abs/1608.07954v3}{{\tt
  http://arxiv.org/abs/1608.07954v3}}].

\bibitem{Bovy:2019uyb}
J.~Bovy, H.~W. Leung, J.~A.~S. Hunt, J.~T. Mackereth, D.~A. Garcia-Hernandez,
  and A.~Roman-Lopes, {\it Life in the fast lane: a direct view of the
  dynamics, formation, and evolution of the milky way's bar},  {\em \mnras}
  {\bf 490} (May, 2019) 4740--4747,
  [\href{http://arxiv.org/abs/http://arxiv.org/abs/1905.11404v2}{{\tt
  http://arxiv.org/abs/1905.11404v2}}].

\bibitem{Wegg2013}
C.~Wegg and O.~Gerhard, {\it Mapping the three-dimensional density of the
  galactic bulge with vvv red clump stars},  {\em \mnras} {\bf 435} (Nov.,
  2013) 1874--1887, [\href{http://arxiv.org/abs/1308.0593}{{\tt
  arXiv:1308.0593}}].

\bibitem{Cox2002}
D.~P. Cox and G.~C. G{\'o}mez, {\it Analytical expressions for spiral arm
  gravitational potential and density},  {\em \apjs} {\bf 142} (Oct., 2002)
  261, [\href{http://arxiv.org/abs/astro-ph/0207635}{{\tt astro-ph/0207635}}].

\bibitem{Miville-Deschenes2016}
M.-A. Miville-Desch\^{e}nes, N.~Murray, and E.~J. Lee, {\it Physical properties
  of molecular clouds for the entire milky way disk},  {\em \apj} {\bf 834}
  (dec, 2016) 57,
  [\href{http://arxiv.org/abs/http://arxiv.org/abs/1610.05918v2}{{\tt
  http://arxiv.org/abs/1610.05918v2}}].

\bibitem{Vasiliev2018}
E.~Vasiliev, {\it Proper motions and dynamics of the milky way globular cluster
  system from gaia dr2},  {\em \mnras} {\bf 484} (jan, 2018) 2832--2850,
  [\href{http://arxiv.org/abs/http://arxiv.org/abs/1807.09775v1}{{\tt
  http://arxiv.org/abs/1807.09775v1}}].

\bibitem{Erkal2016}
D.~Erkal, V.~Belokurov, J.~Bovy, and J.~L. Sanders, {\it The number and size of
  subhalo-induced gaps in stellar streams},  {\em \mnras} {\bf 463} (aug, 2016)
  102--119, [\href{http://arxiv.org/abs/1606.04946}{{\tt arXiv:1606.04946}}].

\bibitem{Springel2008}
V.~Springel, J.~Wang, M.~Vogelsberger, A.~Ludlow, A.~Jenkins, A.~Helmi, J.~F.
  Navarro, C.~S. Frenk, and S.~D. White, {\it The aquarius project: the
  subhaloes of galactic haloes},  {\em \mnras} {\bf 391} (Dec., 2008)
  1685--1711, [\href{http://arxiv.org/abs/0809.0898}{{\tt arXiv:0809.0898}}].

\bibitem{DOnghia2010}
E.~D'Onghia, V.~Springel, L.~Hernquist, and D.~Keres, {\it Substructure
  depletion in the milky way halo by the disk},  {\em \apj} {\bf 709} (jan,
  2010) 1138.

\bibitem{Sawala2016}
T.~Sawala, P.~Pihajoki, P.~H. Johansson, C.~S. Frenk, J.~F. Navarro, K.~A.
  Oman, and S.~D.~M. White, {\it Shaken and stirred: The milky way's dark
  substructures},  {\em \mnras} {\bf 467} (feb, 2017) 4383--4400,
  [\href{http://arxiv.org/abs/1609.01718}{{\tt arXiv:1609.01718}}].

\bibitem{GarrisonKimmel19a}
S.~{Garrison-Kimmel}, P.~F. {Hopkins}, A.~{Wetzel}, J.~S. {Bullock},
  M.~{Boylan-Kolchin}, D.~{Kere{\v{s}}}, C.-A. {Faucher-Gigu{\`e}re},
  K.~{El-Badry}, A.~{Lamberts}, E.~{Quataert}, and R.~{Sand erson}, {\it {The
  Local Group on FIRE: dwarf galaxy populations across a suite of hydrodynamic
  simulations}},  {\em \mnras} {\bf 487} (Jul, 2019) 1380--1399.

\bibitem{Webb2020}
J.~J. {Webb} and J.~{Bovy}, {\it {High-resolution simulations of dark matter
  subhalo disruption in a Milky-Way-like tidal field}},  {\em \mnras} {\bf 499}
  (Nov., 2020) 116--128, [\href{http://arxiv.org/abs/2006.06695}{{\tt
  arXiv:2006.06695}}].

\bibitem{Nadler21a}
E.~O. {Nadler}, A.~{Drlica-Wagner}, K.~{Bechtol}, S.~{Mau}, R.~H. {Wechsler},
  V.~{Gluscevic}, K.~{Boddy}, A.~B. {Pace}, T.~S. {Li}, M.~{McNanna}, A.~H.
  {Riley}, J.~{Garc{\'\i}a-Bellido}, Y.~Y. {Mao}, G.~{Green}, D.~L. {Burke},
  A.~{Peter}, B.~{Jain}, T.~M.~C. {Abbott}, M.~{Aguena}, S.~{Allam},
  J.~{Annis}, S.~{Avila}, D.~{Brooks}, M.~{Carrasco Kind}, J.~{Carretero},
  M.~{Costanzi}, L.~N. {da Costa}, J.~{De Vicente}, S.~{Desai}, H.~T. {Diehl},
  P.~{Doel}, S.~{Everett}, A.~E. {Evrard}, B.~{Flaugher}, J.~{Frieman}, D.~W.
  {Gerdes}, D.~{Gruen}, R.~A. {Gruendl}, J.~{Gschwend}, G.~{Gutierrez}, S.~R.
  {Hinton}, K.~{Honscheid}, D.~{Huterer}, D.~J. {James}, E.~{Krause},
  K.~{Kuehn}, N.~{Kuropatkin}, O.~{Lahav}, M.~A.~G. {Maia}, J.~L. {Marshall},
  F.~{Menanteau}, R.~{Miquel}, A.~{Palmese}, F.~{Paz-Chinch{\'o}n}, A.~A.
  {Plazas}, A.~K. {Romer}, E.~{Sanchez}, V.~{Scarpine}, S.~{Serrano},
  I.~{Sevilla-Noarbe}, M.~{Smith}, M.~{Soares-Santos}, E.~{Suchyta}, M.~E.~C.
  {Swanson}, G.~{Tarle}, D.~L. {Tucker}, A.~R. {Walker}, W.~{Wester}, and {DES
  Collaboration}, {\it {Constraints on Dark Matter Properties from Observations
  of Milky Way Satellite Galaxies}},  {\em \prl} {\bf 126} (Mar., 2021) 091101,
  [\href{http://arxiv.org/abs/2008.00022}{{\tt arXiv:2008.00022}}].

\bibitem{Lovell2013}
M.~R. Lovell, C.~S. Frenk, V.~R. Eke, A.~Jenkins, L.~Gao, and T.~Theuns, {\it
  The properties of warm dark matter haloes},  {\em \mnras} {\bf 439} (feb,
  2014) 300--317, [\href{http://arxiv.org/abs/1308.1399v}{{\tt
  arXiv:1308.1399v}}].

\bibitem{Viel2005}
M.~Viel, J.~Lesgourgues, M.~G. Haehnelt, S.~Matarrese, and A.~Riotto, {\it
  Constraining warm dark matter candidates including sterile neutrinos and
  light gravitinos with wmap and the lyman-$\alpha$ forest},  {\em \prd} {\bf
  71} (mar, 2005) 063534, [\href{http://arxiv.org/abs/astro-ph/0501562}{{\tt
  astro-ph/0501562}}].

\bibitem{Schneider12a}
A.~{Schneider}, R.~E. {Smith}, A.~V. {Macci{\`o}}, and B.~{Moore}, {\it
  {Non-linear evolution of cosmological structures in warm dark matter
  models}},  {\em \mnras} {\bf 424} (July, 2012) 684--698,
  [\href{http://arxiv.org/abs/1112.0330}{{\tt arXiv:1112.0330}}].

\bibitem{Onions12a}
J.~{Onions}, A.~{Knebe}, F.~R. {Pearce}, S.~I. {Muldrew}, H.~{Lux}, S.~R.
  {Knollmann}, Y.~{Ascasibar}, P.~{Behroozi}, P.~{Elahi}, J.~{Han},
  M.~{Maciejewski}, M.~E. {Merch{\'a}n}, M.~{Neyrinck}, A.~N. {Ruiz}, M.~A.
  {Sgr{\'o}}, V.~{Springel}, and D.~{Tweed}, {\it {Subhaloes going Notts: the
  subhalo-finder comparison project}},  {\em \mnras} {\bf 423} (June, 2012)
  1200--1214, [\href{http://arxiv.org/abs/1203.3695}{{\tt arXiv:1203.3695}}].

\bibitem{vbd18a}
F.~C. {van den Bosch}, G.~{Ogiya}, O.~{Hahn}, and A.~{Burkert}, {\it
  {Disruption of dark matter substructure: fact or fiction?}},  {\em \mnras}
  {\bf 474} (Mar., 2018) 3043--3066,
  [\href{http://arxiv.org/abs/1711.05276}{{\tt arXiv:1711.05276}}].

\bibitem{Lovell21a}
M.~R. {Lovell}, M.~{Cautun}, C.~S. {Frenk}, W.~A. {Hellwing}, and O.~{Newton},
  {\it {The spatial distribution of Milky Way satellites, gaps in streams and
  the nature of dark matter}},  {\em arXiv e-prints} (Apr., 2021)
  arXiv:2104.03322, [\href{http://arxiv.org/abs/2104.03322}{{\tt
  arXiv:2104.03322}}].

\bibitem{Schutz20a}
K.~{Schutz}, {\it {Subhalo mass function and ultralight bosonic dark matter}},
  {\em \prd} {\bf 101} (June, 2020) 123026,
  [\href{http://arxiv.org/abs/2001.05503}{{\tt arXiv:2001.05503}}].

\bibitem{Du17a}
X.~{Du}, C.~{Behrens}, and J.~C. {Niemeyer}, {\it {Substructure of fuzzy dark
  matter haloes}},  {\em \mnras} {\bf 465} (Feb., 2017) 941--951,
  [\href{http://arxiv.org/abs/1608.02575}{{\tt arXiv:1608.02575}}].

\bibitem{galacticus}
A.~J. {Benson}, {\it {G ALACTICUS: A semi-analytic model of galaxy formation}},
   {\em New Astronomy} {\bf 17} (Feb., 2012) 175--197,
  [\href{http://arxiv.org/abs/1008.1786}{{\tt arXiv:1008.1786}}].

\bibitem{Amorisco18a}
N.~C. {Amorisco} and A.~{Loeb}, {\it {First constraints on Fuzzy Dark Matter
  from the dynamics of stellar streams in the Milky Way}},  {\em arXiv
  e-prints} (Aug., 2018) arXiv:1808.00464,
  [\href{http://arxiv.org/abs/1808.00464}{{\tt arXiv:1808.00464}}].

\bibitem{corner}
D.~{Foreman-Mackey}, {\it {corner.py: Scatterplot matrices in Python}},  {\em
  The Journal of Open Source Software} {\bf 1} (June, 2016) 24.

\bibitem{Viel13a}
M.~{Viel}, G.~D. {Becker}, J.~S. {Bolton}, and M.~G. {Haehnelt}, {\it {Warm
  dark matter as a solution to the small scale crisis: New constraints from
  high redshift Lyman-{\ensuremath{\alpha}} forest data}},  {\em \prd} {\bf 88}
  (Aug., 2013) 043502, [\href{http://arxiv.org/abs/1306.2314}{{\tt
  arXiv:1306.2314}}].

\bibitem{Irsic17a}
V.~{Ir{\v{s}}i{\v{c}}}, M.~{Viel}, M.~G. {Haehnelt}, J.~S. {Bolton},
  S.~{Cristiani}, G.~D. {Becker}, V.~{D'Odorico}, G.~{Cupani}, T.-S. {Kim},
  T.~A.~M. {Berg}, S.~{L{\'o}pez}, S.~{Ellison}, L.~{Christensen}, K.~D.
  {Denney}, and G.~{Worseck}, {\it {New constraints on the free-streaming of
  warm dark matter from intermediate and small scale
  Lyman-{\ensuremath{\alpha}} forest data}},  {\em \prd} {\bf 96} (July, 2017)
  023522, [\href{http://arxiv.org/abs/1702.01764}{{\tt arXiv:1702.01764}}].

\bibitem{Hsueh20a}
J.~W. {Hsueh}, W.~{Enzi}, S.~{Vegetti}, M.~W. {Auger}, C.~D. {Fassnacht},
  G.~{Despali}, L.~V.~E. {Koopmans}, and J.~P. {McKean}, {\it {SHARP - VII. New
  constraints on the dark matter free-streaming properties and substructure
  abundance from gravitationally lensed quasars}},  {\em \mnras} {\bf 492}
  (Mar., 2020) 3047--3059, [\href{http://arxiv.org/abs/1905.04182}{{\tt
  arXiv:1905.04182}}].

\bibitem{Nadler21b}
E.~O. {Nadler}, S.~{Birrer}, D.~{Gilman}, R.~H. {Wechsler}, X.~{Du},
  A.~{Benson}, A.~M. {Nierenberg}, and T.~{Treu}, {\it {Dark Matter Constraints
  from a Unified Analysis of Strong Gravitational Lenses and Milky Way
  Satellite Galaxies}},  {\em arXiv e-prints} (Jan., 2021) arXiv:2101.07810,
  [\href{http://arxiv.org/abs/2101.07810}{{\tt arXiv:2101.07810}}].

\bibitem{Enzi21a}
W.~{Enzi}, R.~{Murgia}, O.~{Newton}, S.~{Vegetti}, C.~{Frenk}, M.~{Viel},
  M.~{Cautun}, C.~D. {Fassnacht}, M.~{Auger}, G.~{Despali}, J.~{McKean},
  L.~V.~E. {Koopmans}, and M.~{Lovell}, {\it {Joint constraints on thermal
  relic dark matter from strong gravitational lensing, the
  Lyman-{\ensuremath{\alpha}} forest, and Milky Way satellites}},  {\em \mnras}
  (July, 2021) [\href{http://arxiv.org/abs/2010.13802}{{\tt
  arXiv:2010.13802}}].

\bibitem{Gilman20b}
D.~{Gilman}, X.~{Du}, A.~{Benson}, S.~{Birrer}, A.~{Nierenberg}, and T.~{Treu},
  {\it {Constraints on the mass-concentration relation of cold dark matter
  halos with 11 strong gravitational lenses}},  {\em \mnras} {\bf 492} (Feb.,
  2020) L12--L16, [\href{http://arxiv.org/abs/1909.02573}{{\tt
  arXiv:1909.02573}}].

\bibitem{Gilman21a}
D.~{Gilman}, J.~{Bovy}, T.~{Treu}, A.~{Nierenberg}, S.~{Birrer}, A.~{Benson},
  and O.~{Sameie}, {\it {Strong lensing signatures of self-interacting dark
  matter in low-mass halos}},  {\em arXiv e-prints} (May, 2021)
  arXiv:2105.05259, [\href{http://arxiv.org/abs/2105.05259}{{\tt
  arXiv:2105.05259}}].

\bibitem{Vegetti10a}
S.~{Vegetti}, L.~V.~E. {Koopmans}, A.~{Bolton}, T.~{Treu}, and R.~{Gavazzi},
  {\it {Detection of a dark substructure through gravitational imaging}},  {\em
  \mnras} {\bf 408} (Nov., 2010) 1969--1981,
  [\href{http://arxiv.org/abs/0910.0760}{{\tt arXiv:0910.0760}}].

\bibitem{Boddy16a}
K.~K. {Boddy}, M.~{Kaplinghat}, A.~{Kwa}, and A.~H.~G. {Peter}, {\it {Hidden
  sector hydrogen as dark matter: Small-scale structure formation predictions
  and the importance of hyperfine interactions}},  {\em \prd} {\bf 94} (Dec.,
  2016) 123017, [\href{http://arxiv.org/abs/1609.03592}{{\tt
  arXiv:1609.03592}}].

\bibitem{Arvanitaki:2019rax}
A.~Arvanitaki, S.~Dimopoulos, M.~Galanis, L.~Lehner, J.~O. Thompson, and
  K.~Van~Tilburg, {\it {Large-misalignment mechanism for the formation of
  compact axion structures: Signatures from the QCD axion to fuzzy dark
  matter}},  {\em Phys. Rev. D} {\bf 101} (2020), no.~8 083014,
  [\href{http://arxiv.org/abs/1909.11665}{{\tt arXiv:1909.11665}}].

\bibitem{Armengaud17a}
E.~{Armengaud}, N.~{Palanque-Delabrouille}, C.~{Y{\`e}che}, D.~J.~E. {Marsh},
  and J.~{Baur}, {\it {Constraining the mass of light bosonic dark matter using
  SDSS Lyman-{\ensuremath{\alpha}} forest}},  {\em \mnras} {\bf 471} (Nov.,
  2017) 4606--4614, [\href{http://arxiv.org/abs/1703.09126}{{\tt
  arXiv:1703.09126}}].

\bibitem{Irsic17b}
V.~{Ir{\v{s}}i{\v{c}}}, M.~{Viel}, M.~G. {Haehnelt}, J.~S. {Bolton}, and G.~D.
  {Becker}, {\it {First Constraints on Fuzzy Dark Matter from
  Lyman-{\ensuremath{\alpha}} Forest Data and Hydrodynamical Simulations}},
  {\em \prl} {\bf 119} (July, 2017) 031302,
  [\href{http://arxiv.org/abs/1703.04683}{{\tt arXiv:1703.04683}}].

\bibitem{Kobayashi17a}
T.~{Kobayashi}, R.~{Murgia}, A.~{De Simone}, V.~{Ir{\v{s}}i{\v{c}}}, and
  M.~{Viel}, {\it {Lyman-{\ensuremath{\alpha}} constraints on ultralight scalar
  dark matter: Implications for the early and late universe}},  {\em \prd} {\bf
  96} (Dec., 2017) 123514, [\href{http://arxiv.org/abs/1708.00015}{{\tt
  arXiv:1708.00015}}].

\bibitem{Chan20a}
J.~H.~H. {Chan}, H.-Y. {Schive}, S.-K. {Wong}, T.~{Chiueh}, and
  T.~{Broadhurst}, {\it {Multiple Images and Flux Ratio Anomaly of Fuzzy
  Gravitational Lenses}},  {\em \prl} {\bf 125} (Sept., 2020) 111102,
  [\href{http://arxiv.org/abs/2002.10473}{{\tt arXiv:2002.10473}}].

\bibitem{Hezaveh2016}
Y.~D. Hezaveh, N.~Dalal, D.~P. Marrone, Y.-Y. Mao, W.~Morningstar, D.~Wen,
  R.~D. Blandford, J.~E. Carlstrom, C.~D. Fassnacht, G.~P. Holder, A.~Kemball,
  P.~J. Marshall, N.~Murray, L.~P. Levasseur, J.~D. Vieira, and R.~H. Wechsler,
  {\it Detection of lensing substructure using alma observations of the dusty
  galaxy sdp.81},  {\em \apj} {\bf 823} (May, 2016) 37,
  [\href{http://arxiv.org/abs/http://arxiv.org/abs/1601.01388v1}{{\tt
  http://arxiv.org/abs/1601.01388v1}}].

\bibitem{Jungman:1995df}
G.~Jungman, M.~Kamionkowski, and K.~Griest, {\it Supersymmetric dark matter},
  {\em Physics Reports} {\bf 267} (mar, 1996) 195--373,
  [\href{http://arxiv.org/abs/hep-ph/9506380}{{\tt hep-ph/9506380}}].

\bibitem{Bertone05}
G.~Bertone, D.~Hooper, and J.~Silk, {\it {Particle dark matter: Evidence,
  candidates and constraints}},  {\em Phys. Rept.} {\bf 405} (2005) 279--390,
  [\href{http://arxiv.org/abs/hep-ph/0404175}{{\tt hep-ph/0404175}}].

\bibitem{Carr16a}
B.~{Carr}, F.~{K{\"u}hnel}, and M.~{Sandstad}, {\it {Primordial black holes as
  dark matter}},  {\em \prd} {\bf 94} (Oct., 2016) 083504,
  [\href{http://arxiv.org/abs/1607.06077}{{\tt arXiv:1607.06077}}].

\bibitem{Preskill83a}
J.~{Preskill}, M.~B. {Wise}, and F.~{Wilczek}, {\it {Cosmology of the invisible
  axion}},  {\em Physics Letters B} {\bf 120} (Jan., 1983) 127--132.

\bibitem{Abbott83a}
L.~F. {Abbott} and P.~{Sikivie}, {\it {A cosmological bound on the invisible
  axion}},  {\em Physics Letters B} {\bf 120} (Jan., 1983) 133--136.

\bibitem{Dine83a}
M.~{Dine} and W.~{Fischler}, {\it {The not-so-harmless axion}},  {\em Physics
  Letters B} {\bf 120} (Jan., 1983) 137--141.

\bibitem{Dodelson1994}
S.~Dodelson and L.~M. Widrow, {\it Sterile neutrinos as dark matter},  {\em
  Physical Review Letters} {\bf 72} (jan, 1994) 17.

\bibitem{Huo18a}
R.~{Huo}, M.~{Kaplinghat}, Z.~{Pan}, and H.-B. {Yu}, {\it {Signatures of
  self-interacting dark matter in the matter power spectrum and the CMB}},
  {\em Physics Letters B} {\bf 783} (Aug., 2018) 76--81,
  [\href{http://arxiv.org/abs/1709.09717}{{\tt arXiv:1709.09717}}].

\end{thebibliography}\endgroup

\end{document}